\documentclass[preprint,preprintnumbers,aps]{revtex4}

\usepackage[dvips]{graphicx}
\usepackage{graphicx}
\usepackage{amsfonts}
\usepackage{bm}
\usepackage{amsmath}
\usepackage{amssymb}
\usepackage{color}
\usepackage[all]{xy}
\usepackage{mathtools}

\usepackage{booktabs}

\usepackage{float}
\usepackage{subcaption}

\def\be{\begin{equation}}
\def\ee{\end{equation}}
\def\bea{\begin{eqnarray}}
\def\eea{\end{eqnarray}}

\newcommand{\der}[2]{\frac{d #1}{d #2}}

\newcommand{\derp}[2]{\frac{\partial #1}{\partial #2}}
\newcommand{\sch}{Schwarzschild}

\usepackage[dvipsnames]{xcolor}

\begin{document}

\title{Spherically symmetric collapse: Initial configurations}

\author{Elly Bayona$^1$, and Hernando Quevedo$^{1,2,3}$, Miguel Alcubierre$^1$,}
\affiliation{$^1$Instituto de Ciencias Nucleares, Universidad Nacional Aut\'onoma de M\'exico, México.}
\affiliation{$^2$Dipartimento di Fisica and Icra, Universit\`a di Roma “La Sapienza”, Roma, Italy.}
\affiliation{$^3$ Al-Farabi Kazakh National University, Al-Farabi av. 71, 050040 Almaty, Kazakhstan.}

\date{\today}


\begin{abstract}

The initial state of the spherical gravitational collapse in general relativity has been studied with different methods, especially by using {\it a priori} given equations of state that describe the matter as a perfect fluid. We propose an alternative approach, in which the energy density of the perfect fluid is given as a polynomial function of the radial coordinate that is well-behaved everywhere inside the fluid. We then solve the corresponding differential equations, including the Tolman-Oppenheimer-Volkoff equilibrium condition, using a fourth-order Runge-Kutta method and obtain a consistent model with a central perfect-fluid core surrounded by dust. We analyze the Hamiltonian constraint, the mass-to-radius relation, the boundary and physical conditions, and the stability and convergence properties of the numerical solutions. The energy density and pressure of the resulting matter distribution satisfy the standard physical conditions. The model is also consistent with the Buchdahl limit and the speed of sound conditions, even by using realistic values of compact astrophysical objects such as neutron stars.


\end{abstract}


\maketitle


\section{Introduction}
\label{sec:int}

The collapse of a mass distribution is of particular importance for analyzing the properties of the gravitational field under extreme conditions. 
Roughly speaking, the study of the collapse can be divided into two parts, namely, the initial configuration and the dynamics involved in the evolution of that initial configuration. In this work, we will focus on the initial configurations only and leave the dynamics of the collapse for future work. We limit ourselves in this work to the study of spherically symmetric configurations.

Birkhoff's theorem states that the only spherically symmetric vacuum solution of Einstein's equations is the Schwarzschild metric \cite{thorne2000gravitation}. This facilitates the study of black hole solutions in Einstein's gravity. The case of interior spherically symmetric solutions is more complicated. In general, such interior gravitational fields are assumed to be generated by perfect fluids, and the usual approach to obtain such solutions is to specify {\it a priori} a particular equation of state (EoS), which is commonly assumed as corresponding to barotropic or polytropic fluids. Although there exist in the literature more than a hundred solutions of this type, unfortunately only a few of them satisfy the most elementary physical conditions \cite{delgaty1998physical}.
For instance, the celebrated interior Schwarzschild solution \cite{stephani2004relativity} with a constant energy density leads to a sound speed that violates the light speed limit. In our opinion, most of the problems related to spherically symmetric perfect-fluid solutions are due to the choice of specific EoS's with the aim of obtaining exact analytic solutions. In this work, we propose a different approach.

We start with the energy density of the fluid, which we fix as a polynomial function of the radial coordinate with arbitrary coefficients. The main advantage of this approach is that we can fix {\it a priori} the behavior of the energy density by choosing the values of the free coefficients appropriately. In particular, we can demand that the energy density is a decreasing function with the maximum value at the center of the sphere and with a vanishing value at the radius of the sphere. 

It turns out that this simple assumption allows us to construct a model that can be completely controlled with the values of the coefficients of the energy density and the value of the pressure at the center of the sphere. As a result, we obtain several models that can be used to describe the gravitational field of a spherically symmetric mass distribution, satisfying the main physical conditions, namely, the energy density and the pressure are free of singularities and vanish at the exterior of the star, the sound speed satisfies the light speed limit everywhere inside the fluid, and the compactness of the mass distribution satisfies the Buchdahl limit. In addition, we show that the numerical approach used to construct the model satisfies the converge conditions expected for a numerical solution.

The model we will present consists essentially of a central core surrounded by a layer of dust, the thickness of which can be modified by changing the values of the parameters determining the model. Moreover, we show that the parameters can be fixed using the physical quantities that characterize realistic stars, such as neutron stars. 

This work is organized as follows. In Sec. \ref{sec:PerfFluid}, we review the main equations used in general relativity to describe a spherically symmetric perfect fluid and present our proposal for the energy density profile in terms of the radial coordinate. In Sec. \ref{sec:EqCond}, we describe the boundary and physical conditions imposed for the model to be physically meaningful. In Sec. \ref{sec:numSol} we present the numerical solutions that are obtained by solving numerically the Tolman-Oppenheimer-Volkoff (TOV) equation, and describe the details of the main physical quantities that can be obtained from the numerical solutions. In Sec. \ref{sec:NS} we show that the model can be used with the physical parameters that characterize canonical neutron stars. Then, in Sec. \ref{sec:StaCon}, we test the stability and convergence properties of the model. Finally, in Sec.\ref{sec:con} we review our results and comment on possible applications of our model.


\section{Perfect fluids }
\label{sec:PerfFluid}

Consider a perfect fluid described with mass-energy density $\rho$ and pressure $p$, as measured in its rest frame. The corresponding  stress-energy-momentum tensor is given by
\begin{equation}
T_{\mu\nu}=(\rho+p)u_\mu u_\nu +pg_{\mu\nu} \; ,
\end{equation}
where $u^\mu=g^{\mu\nu}u_\nu$ are the components of the four-velocity of the fluid element. The main physical property of a perfect fluid is that it corresponds to an idealized matter model because it does not consider physical quantities such as shear stresses, viscosity, or heat conduction, which are characteristics of real fluids.  

Consider now a static and spherically symmetric spacetime represented by the general line element
\begin{equation}
ds^2 =  -e^{2\phi} dt^2 +e^{2\psi}dr^2 +r^2(d\theta^2+\sin^2\theta d\phi^2) \; ,
\end{equation}
where $\phi$ and $\psi$ are functions of the radial coordinate $r$ only. In this case, $u^\mu=e^{\phi(r)}(\partial_t)^\mu$, so that the four-velocity is normalized as  $u^\mu u_\mu=-1$. Then, the independent components of Einstein's equations can be written as follows
\begin{align}
8\pi\rho &= \frac{2}{r} e^{-2\psi}\psi_r+\frac{1-e^{-2\psi}}{r^2} \; ,\label{1Ee}\\
8\pi p &= \frac{2}{r} e^{-2\psi}\phi_r+\frac{e^{-2\psi}-1}{r^2} \; , \label{2Ee}\\
8\pi p&=e^{-2\psi}\left[ (\phi_r)^2 + \phi_{rr} - \phi_r \psi_r + \frac{\phi_r - \psi_r}{r} \right] \; , \label{3Ee}
\end{align}
where $\phi_r = \frac{\partial \phi}{\partial r} $, etc. From~\eqref{1Ee} we obtain
\begin{equation}
A(r) \coloneq e^{2\psi(r)} = \left( 1 - \frac{2m(r)}{r} \right)^{-1}, \qquad\qquad m(r) \coloneq 4 \pi \int_0^r \rho(r') r'^2 dr' \; . \label{massFun}
\end{equation}
where we have introduced the mass function $m(r)$. Then, from  \eqref{2Ee} we have that
\begin{equation}
\frac{d\phi}{dr} = \frac{m(r)+4\pi r^3p(r)}{r(r-2m(r))} \; . \label{phiEq}
\end{equation}
Finally, using Eqs.\eqref{1Ee},  \eqref{2Ee} and \eqref{3Ee}, we obtain 
\begin{equation}
p_r = -(p+\rho)\phi_r\label{MovEq} \; ,
\end{equation}
which is equivalent to the conservation law $\nabla^\mu T_{\mu\nu}=0$ and can be represented as the Tolman-Oppenheimer-Volkoff (TOV) equation,
\begin{equation}
\frac{dp}{dr} = -\Big( p(r) + \rho(r) \Big) \:
\frac{m(r)+4\pi r^3p(r)}{r(r-2m(r))} \; . \label{TOVEq}
\end{equation}
For more details, see the review \cite{Sarbach_2021}. 

We see that we have three equations \eqref{massFun}, \eqref{phiEq} and \eqref{TOVEq} with four unknown variables, namely, the two metric functions $\phi(r)$ and  $A(r)=e^{2\psi(r)}$, and the two fluid variables $p(r)$ and $\rho(r)$. To search for solutions to this system of equations it is necessary to add an extra equation, which is usually taken to be an equation of state of the form $p=p(\rho)$. As mentioned above, this approach leads in many cases to solutions that do not satisfy the physical conditions expected for a realistic spherically symmetric mass distribution. Here, we will proceed in a different way.


\subsection{Predetermined density profiles}
\label{sec:den}

The alternative approach we propose in this work consists in giving {\it a priori} the mass-energy density as a function of a radial coordinate in such a way that it satisfies the expected physical conditions, namely, it should be positive and free of divergences in the entire region inside the sphere that contains the mass distribution. Such a profile for the density will guarantee {\it a priori} that the behavior of the energy density corresponds to that of a realistic astrophysical object. In particular, we can fix the density function as a polynomial of the form
\begin{equation}
\rho(r) = \sum_{i=0}^n c_ir^i \; ,
\end{equation}
where $c_i$ represents a series of real constants. For concreteness, we will limit ourselves to the case $n=4$, which will prove to lead to satisfactory results. We assume that the origin of coordinates is located at the center of the object. This fixes the meaning and the value of one of the coefficients $c_i$, namely,  $c_0=\rho(r=0),\equiv \rho_c$. The remaining coefficients will determine the internal structure of the star. For the sake of concreteness, we choose an energy density profile of the form 
\begin{equation}
\rho(r) = \rho_c - c_1r - c_2r^2 - c_3r^3 - c_4r^4 \; . \label{1Rho}
\end{equation}
This expression represents the additional equation we need in order to integrate Einstein's equations.


\section{Boundary and physical conditions}
\label{sec:EqCond}

In order to solve the TOV equations numerically, we will use \eqref{1Rho} instead of the typical polytropic EoS. In order to have a regular solution at the origin we assume the density to be a smooth function at the center of the object. This means that we must choose an even profile for the total energy density, i.e.,
\begin{equation}
\label{2Rho}
\rho(r) =
\begin{cases}
\rho_c - c_2r^2 - c_4r^4 \; , & r<R \\
0 \; , & r>R \; ,
\end{cases}
\end{equation}  
with $R$ the radius of the star, which can be defined as the coordinate point where the total energy density is zero $\rho(r=R)=0$, i.e., 
\begin{equation}\label{rho_cR}
\rho_c = c_4R^4 + c_2R^2\ .
\end{equation}
This implies a relationship between the central total energy density and the radius of the object. Eq. \eqref{2Rho} now allows us to write the mass function $m(r)$ as
\begin{equation}\label{mass}
m(r) =
\begin{cases}
4 \pi r^3 \left[ \dfrac{\rho_c}{3} - \dfrac{c_2}{5}r^2 - \dfrac{c_4}{7}r^4 \right] \; , & r<R \; ,\\
4 \pi R^3 \left[ \dfrac{\rho_c}{3} - \dfrac{c_2}{5}R^2 - \dfrac{c_4}{7}R^4 \right] \equiv M \; , & r>R \; .
\end{cases}
\end{equation} 
Thus, the metric function $A (r)$ can be expressed in the following manner
\begin{equation}\label{Ar}
A(r) = \left(1-\frac{2m(r)}{r}\right)^{-1} =
\begin{cases}
\left(1-8\pi r^2\left[ \dfrac{\rho_c}{3} - \dfrac{c_2r^2}{5}
- \dfrac{c_4r^4}{7} \right] \right)^{-1} \; , &r<R \\
\left( 1 - \dfrac{2M}{r} \right)^{-1} \; , & r>R \; .
\end{cases}
\end{equation}

We will still assume that the perfect fluid is in local thermodynamic equilibrium and is isotropic. As shown in \cite{Sarbach_2021}, this assumption implies that each fluid cell is in thermodynamic equilibrium and the object is at a maximum of entropy. We also assume that each cell has a fixed number $N$ of particles and the volume is given by $V=N/n$, where $n$ refers to the particle density. Subsequently, we have
\begin{equation}\label{de1}
de = - p \: d\left(\frac{1}{\rho_0}\right) =\frac{p}{\rho_0^2} \: d\rho_0 \; .
\end{equation}
Here, {$e=U/M$} is the internal specific energy, and $\rho_0$ is the rest mass energy density. Then, we have two new quantities that describe the state of the fluid and are related to the total energy density through
\begin{equation}
\rho(r) = \rho_0(r)(1 + e(r)) \ . \label{TotRho}
\end{equation}
Calculating the derivative of \eqref{TotRho}, and matching it with \eqref{de1}, we obtain
\begin{equation}\label{derRho0}
\der{\rho_0}{r} = \frac{\rho_0}{[p(r) + \rho(r)]} \: \der{\rho}{r} \ ,
\end{equation}
where the total energy density derivative is clearly given by
\begin{equation}\label{derRho}
\der{\rho}{r} =
\begin{cases}
- 2r(c_2 + 2c_4r^2) \; , &r<R \; ,\\
0 \; , &r>R \; .
\end{cases}
\end{equation}
Therefore, \eqref{TotRho} and \eqref{derRho0} are two additional equations that are needed in order to find the thermodynamic quantities $\rho_0(r)$ and $e(r)$.

In summary, our system of equations contains two metric functions ($\phi(r)$ and $A(r)$, or equivalently $m(r)$), and four thermodynamic quantities ($p(r)$, $\rho(r)$, $\rho_0(r)$, $e(r)$), which should be solved using Eqs.  
\eqref{massFun}, \eqref{phiEq}, \eqref{TOVEq}, \eqref{2Rho}, \eqref{TotRho}, and \eqref{derRho0}. 

Another important physical quantity is the speed of sound inside the fluid, $v_s$, which is  defined as
\begin{equation}
v_s^2 = \derp{p}{\rho} \; . \label{vs}
\end{equation}
For the numerical estimation of this quantity, we integrate the TOV equations to obtain $p=p(r)$ and use the expression (\ref{2Rho}) to evaluate $\rho=\rho(r)$, so that for different values of $r$ we can calculate
\begin{equation}
v_s^2 = \frac{\Delta p}{\Delta \rho} \; ,
\end{equation}
up to the second order. For the sound speed to be physical, we demand that it satisfies the light speed limit, i.e., $v_s^2<1.$


\subsection{Behavior at $r=0$.}
\label{subsec:r0}

From a physical (and numerical) point of view, it is important to examine the behavior of certain quantities and differential equations that may be diverging at the center of the object $r=0$. For example, at $r=0$ the metric function $A(r)$ is finite and tends to 1, as can be seen from the right-hand side of Eq.\eqref{Ar}.  Moreover, by replacing the mass function \eqref{mass} in Eqs.\eqref{TOVEq} and \eqref{2Rho}, and evaluating at $r=0$, one obtains: 
\begin{equation}
\left. \der{p}{r}\right|_{r=0} = 0 \ , \qquad \qquad
\left. \der{\rho_0}{r} \right|_{r=0} = 0 \ .
\end{equation}

Furthermore, it is possible to find the analytical value of the speed of sound at the center of the object. To this end, we first replace the value of the function $m(r)$ from Eq. \eqref{mass} in the TOV equation~\eqref{TOVEq}.  We then use the expression for the derivative of the total energy density~\eqref{2Rho} and, finally, use the chain rule to obtain
\begin{equation}
\left. v_s^2 \right|_{r=0} = \left[ \der{p}{r} \left( \der{\rho}{r} \right)^{-1} \right]_{r=0}
= \frac{2\pi}{c_2} \:
(p_c + \rho_c) \left( p_c + \frac{\rho_c}{3} \right) \; , \label{vscenter}
\end{equation}
which should also satisfy the light-speed limit
\begin{equation}
\frac{2\pi}{c_2} \: (p_c+\rho_c) \left(p_c+\frac{\rho_c}{3} \right)
< 1.
\end{equation}

As a result, the expression for the total energy density \eqref{2Rho} ensures that our system of equations behaves properly at $r=0$.


\subsection{Additional physical conditions}
\label{sec:PhysicalCond}

We will impose two additional physical conditions in order to obtain realistic configurations. The first condition is related to the stability of the object and is known as the Buchdahl limit, which states that the compactness ratio of the object  should be bounded from above, i.e., 
\begin{equation}
\frac{2m(r)}{r} < \frac{8}{9} \; . \label{buch}
\end{equation}
Although this limit was first shown for incompressible spheres \cite{buchdahl1959general}, 
it can be generalized to include a wide class of physically reasonable equations of state \cite{Sarbach_2021}.

Notice that the Buchdahl condition is assumed to obtain stable configurations that could describe the gravitational field of realistic astrophysical bodies. However, if we are interested in configurations that can undergo a gravitational collapse, the Buchdahl limit represents a useful tool to control the beginning of the collapse. In fact, if we assume a compactness ratio such that $2m(r)/r \geq 8/9$, the initial condition for a gravitational collapse is guaranteed.

A consequence of imposing the  Buchdahl limit \eqref{buch} is the fact that the metric function $A(r)$ does not diverge, i.e.,
\begin{equation}
    A(r) = \left( 1 - \frac{2m(r)}{r} \right)^{-1}<9 \ , \label{acond}
\end{equation}
which is a condition that must be taken into account when solving the system of equations that determine the spherical configurations.

The second physical condition that we will impose states that 
\begin{equation}
    e_c = \frac{\rho_c}{\rho_{0_c}} - 1 \geq 0 \ ,
\end{equation}
which means that the specific internal energy must be positive, implying that the central total energy density must be greater than the central rest energy density.


\section{Numerical Solutions}
\label{sec:numSol}

To solve the problem discussed in the previous sections, we should consider two important aspects.  
The first concerns the metric function $\phi(r)$. If we find the radial dependence of the mass function $m(r)$ and the pressure $p(r)$, we can then easily find the solution for $\phi(r)$ by simple integration using \eqref{phiEq}. We will use this procedure to find the explicit behavior of $\phi(r)$.

The second aspect is related to the free parameters of the problem.  To determine the energy density profile \eqref{2Rho}, it is necessary to specify three parameters, namely the central energy density $\rho_c$ and the two coefficients $c_2$ and $c_4$. Furthermore, to solve the TOV differential equations \eqref{TOVEq} and \eqref{derRho0} it is necessary to fix the central pressure $p_c$ and the central internal specific energy $e_c$. Consequently, we need to fix five parameters before solving the problem numerically. It turns out that there are several possibilities that lead to physically reasonable solutions. In Table~\ref{Cases} we present the values of the parameters for the two representative models that we will analyze in detail. 

Finally, as previously mentioned, we intend to utilize the results of the present work as initial configuration for studying the collapse of a spherically symmetric mass distribution, for which we will use the Ollinsphere code that uses the Baumgarte-Shapiro-Shibata-Nakamura (BSSN) formulation~\cite{Baumgarte:1998te,shibata95,alcubierre2008introduction} adapted to spherical symmetry~\cite{alcubierre2011formulations}. This code has been described before for example in~\cite{Alcubierre:2019qnh,Degollado:2020lsa}. Therefore, the numerical code developed here has been adapted to the requirements of the OllinSphere code. To solve the TOV differential equations, we will use a fourth-order Runge-Kutta method~\cite{press1988numerical}.

\begin{table}
    \centering
    \begin{tabular}{|l|r|r|}
    \hline
    \textbf{Parameter} & \textbf{\ Model 1$\ 
 \ $}&\textbf{\ Model 2$\ \ $} \\
        \hline
        \hline
         Central rest energy density $\rho_{c}$ \ & $0.08804$ &$0.14244$\\
         Central internal energy $e_c$ &$0.05$& $0.11$ \\
         Central pressure $p_c$ & $0.01$&$0.02$ \\
         Constant $c_2$ & $0.04179$ &$0.08259$\\
         Constant $c_4$ & $0.083$ &$0.001$\\
         \hline
    \end{tabular}
    \centering
    \caption{Two sets of parameters necessary for integrating the TOV equations. Model 1 leads to a central sound speed $v_s|_{r=0} = 0.76$ and radius $R=0.9$, while for model 2 we obtain  $v_s|_{r=0}=0.92$ and $R=1.3$.}
    \label{Cases}
\end{table}

\vspace{5mm}

\begin{figure}[h]
     \begin{subfigure}[b]{0.49\textwidth}
         \includegraphics[width=\textwidth]{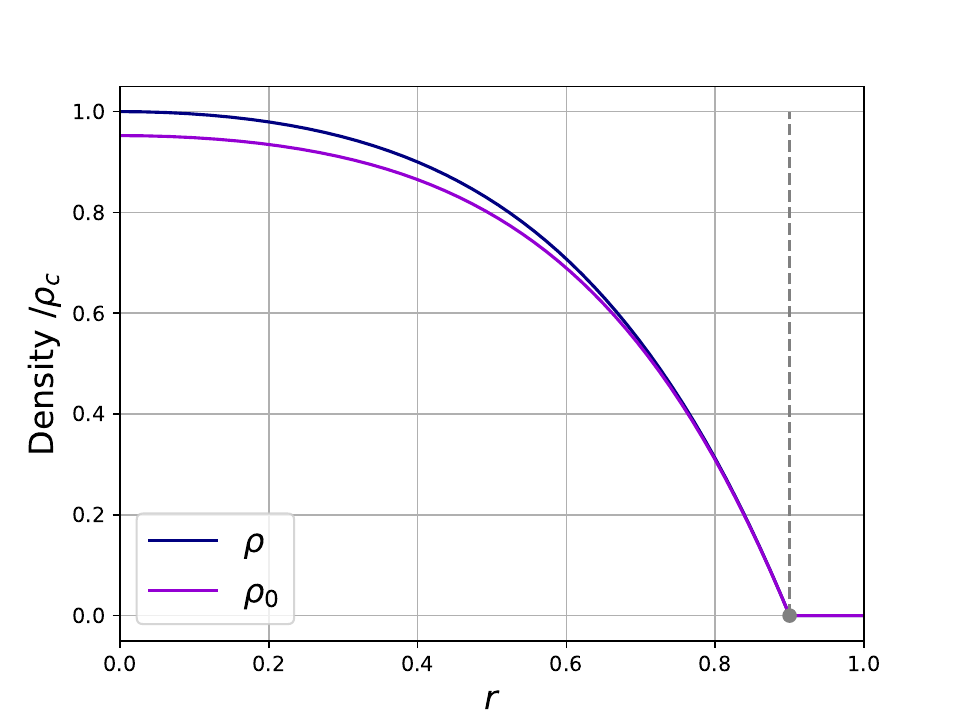}
         \caption{Energy densities for model 1, with radius $R=0.9$.}
         \label{rho0caso077}
     \end{subfigure}
     \hfill
     \begin{subfigure}[b]{0.49\textwidth}
         \includegraphics[width=\textwidth]{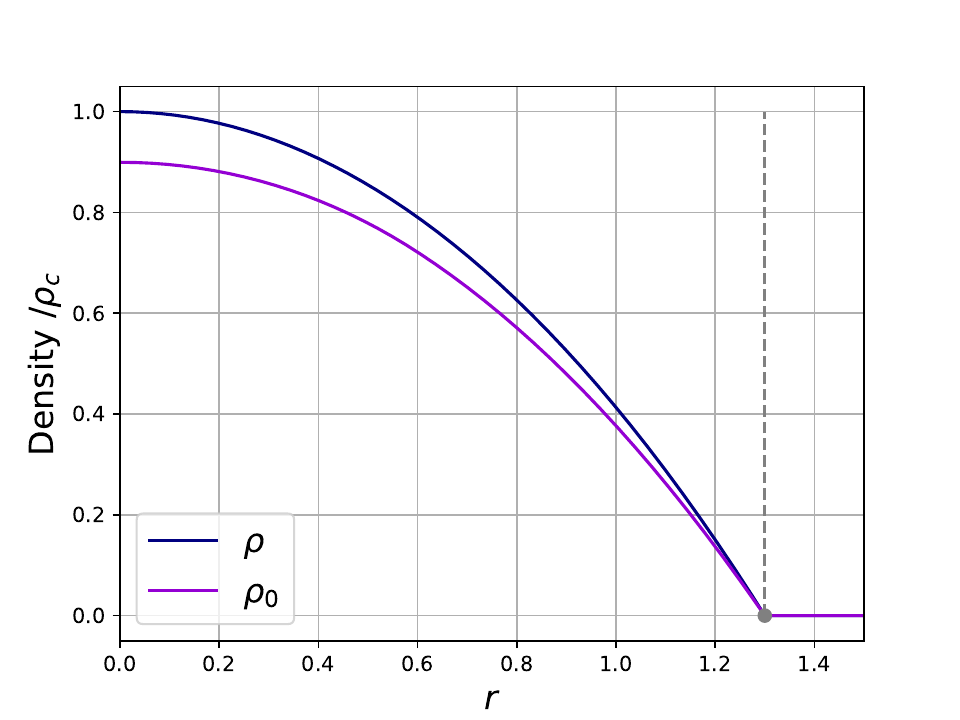}
         \caption{Energy densities for model 2 of, with radius $R=1.3$.}
         \label{rho0caso095}
     \end{subfigure}
\caption{Rest mass and total energy density profiles $\rho_0(r)$ and $\rho(r)$ for the two models presented in table \ref{Cases}. The radius of the object is represented by the dashed grey vertical line.}
\label{rho_plot}
\end{figure}

\begin{figure}[h]
     \begin{subfigure}[b]{0.49\textwidth}
         \centering
         \includegraphics[width=\textwidth]{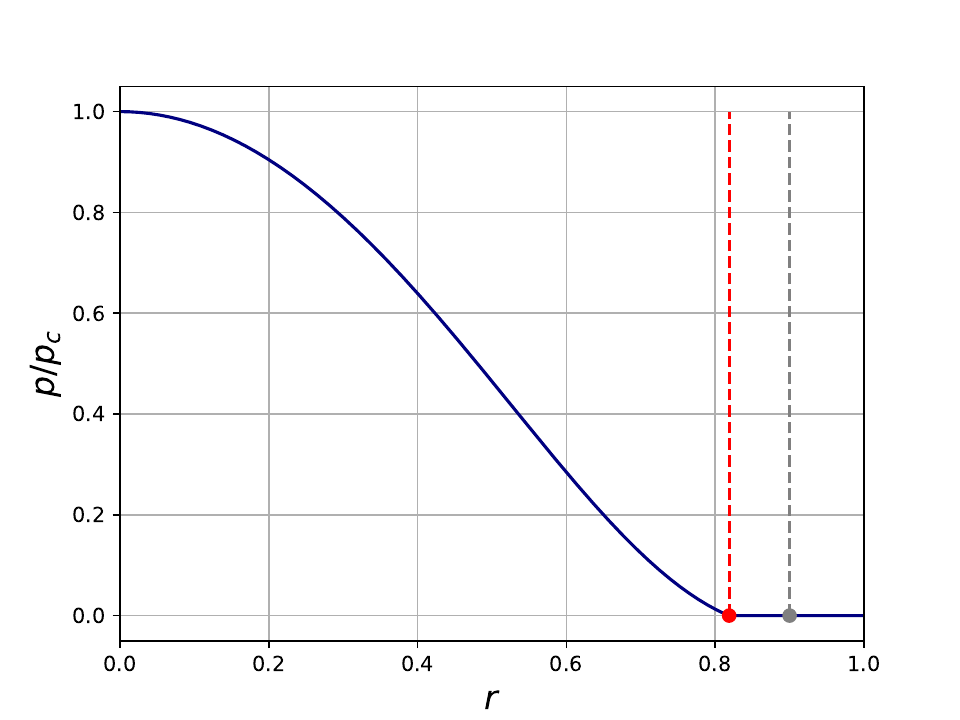}
         \caption{Pressure for model 1 ,with zero-pressure radius $r_{p1}=0.82$.}
         \label{p0caso077}
     \end{subfigure}
     \hfill
     \begin{subfigure}[b]{0.49\textwidth}
         \centering
         \includegraphics[width=\textwidth]{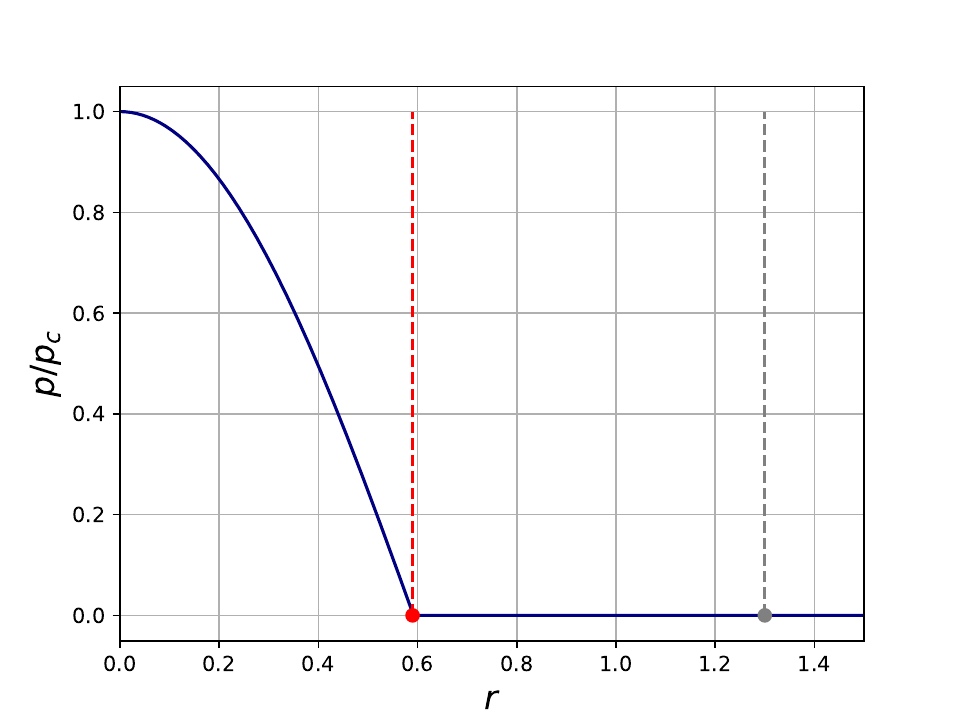}
         \caption{Pressure for model 2, with zero-pressure radius $r_{p2}=0.6$.}
         \label{p0caso095}
     \end{subfigure}
        \caption{Pressure profiles $p(r)$ for the two models presented in table \ref{Cases}. The radius of the object is again represented by dashed grey vertical lines, while the radius at which the pressure becomes zero is denoted by dashed red vertical lines.}
        \label{p0}
\end{figure}

We now present the main results of the numerical integration. In Fig.~\ref{rho_plot}, we plot the profile of the total and rest mass energy densities. In both cases we observe that the energy densities are decreasing functions of the radial coordinate $r$, reaching the zero-value at the radius of the object $r=R$. The pressure profile $p(r)$ is presented in Fig.~\ref{p0}. It turns out that in both cases the pressure becomes zero at a radial distance smaller than the radius of the object. For model 1 we find that the pressure becomes zero at a radius $r_{p1}=0.82$, while for model 2 this happens at $r_{p2}=0.6$. This means that the spherical configuration consists of a central core surrounded by a layer of dust with zero pressure, which extends from $r=r_p$ to $r=R$. In models 1 and 2 the thickness of this layer is $0.1$ and $0.8$ respectively. Interestingly, the dust layer is always present for all the explored choices of parameters that satisfy the physical conditions.

\begin{figure}[h]
     \centering
     \begin{subfigure}[b]{0.49\textwidth}
         \centering
         \includegraphics[width=\textwidth]{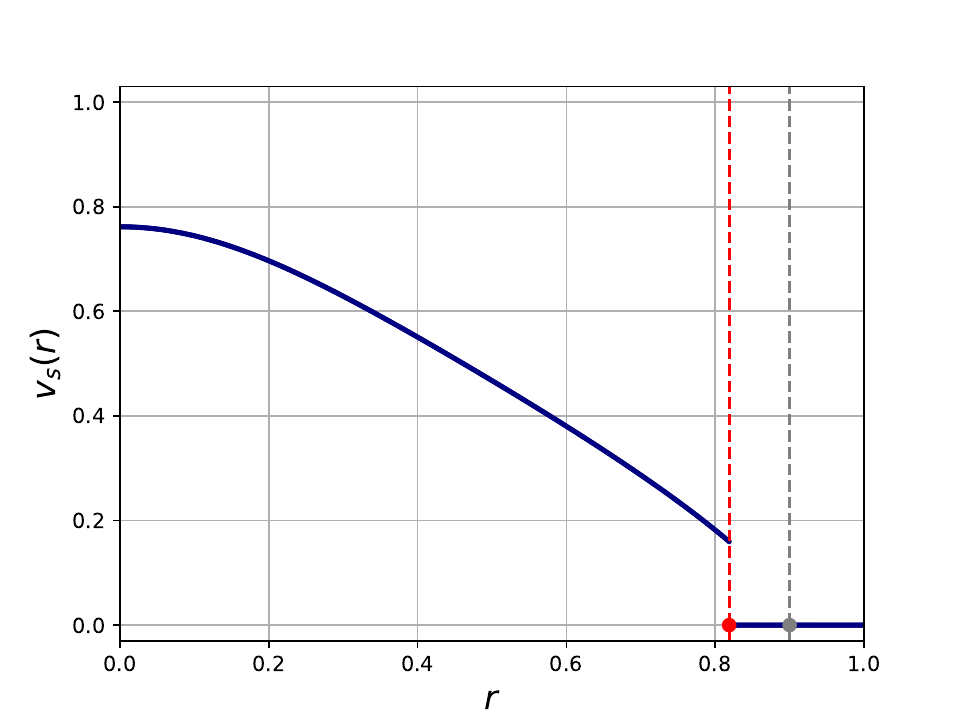}
         \caption{Sound speed for model 1.}
         \label{vs0caso077}
     \end{subfigure}
     \hfill
     \begin{subfigure}[b]{0.49\textwidth}
         \centering
         \includegraphics[width=\textwidth]{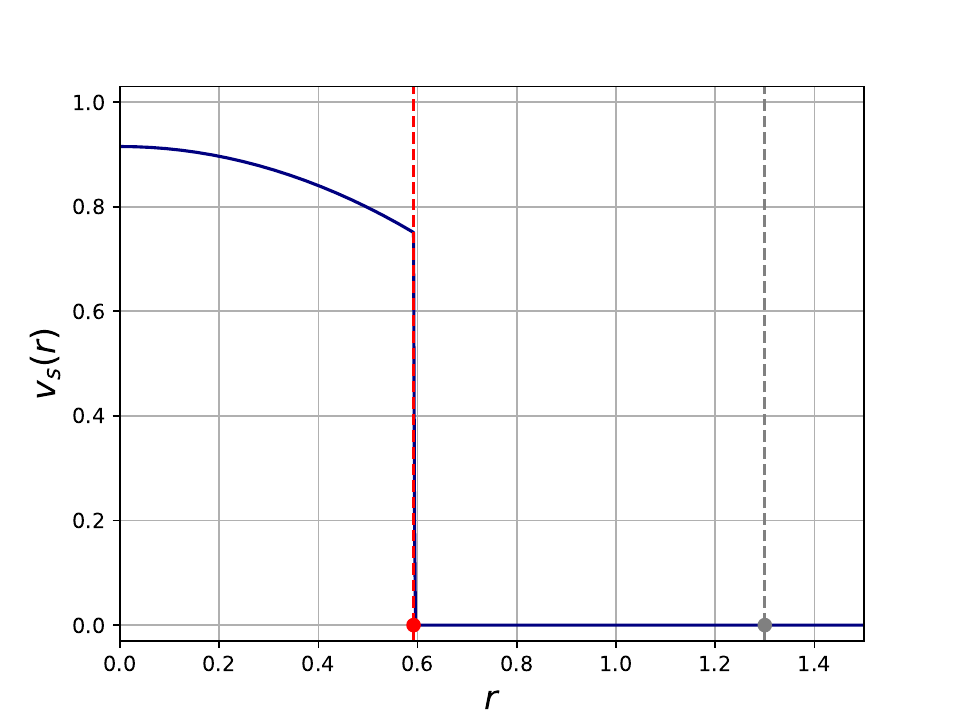}
         \caption{Sound speed for model 2.}
         \label{vs0caso095}
     \end{subfigure}
        \caption{Sound speed $v_s(r)$ for the two models presented in the table \ref{Cases}. The radius of the star is shown as a gray vertical line.}
        \label{vs0}
\end{figure}

The speed of sound for both models is shown in Fig.~\ref{vs0}. The maximum values are $v_s=0.76$ for model 1 and $v_s=0.92$ for model 2 and are reached at the center of the object in both cases. 
Notice that even when the parameters of the Table \ref{Cases} were forced to satisfy the speed of light limit only at the center of the object, this limit is valid for all values of the radius. Moreover, the speed of sound decreases monotonically as expected from a physical viewpoint. Therefore, in both cases, the speed of light limit is completely satisfied. Additionally, note that both models exhibit a discontinuity at the point where the pressure vanishes.

\begin{figure}[h]
     \centering
     \begin{subfigure}[b]{0.49\textwidth}
         \centering
         \includegraphics[width=\textwidth]{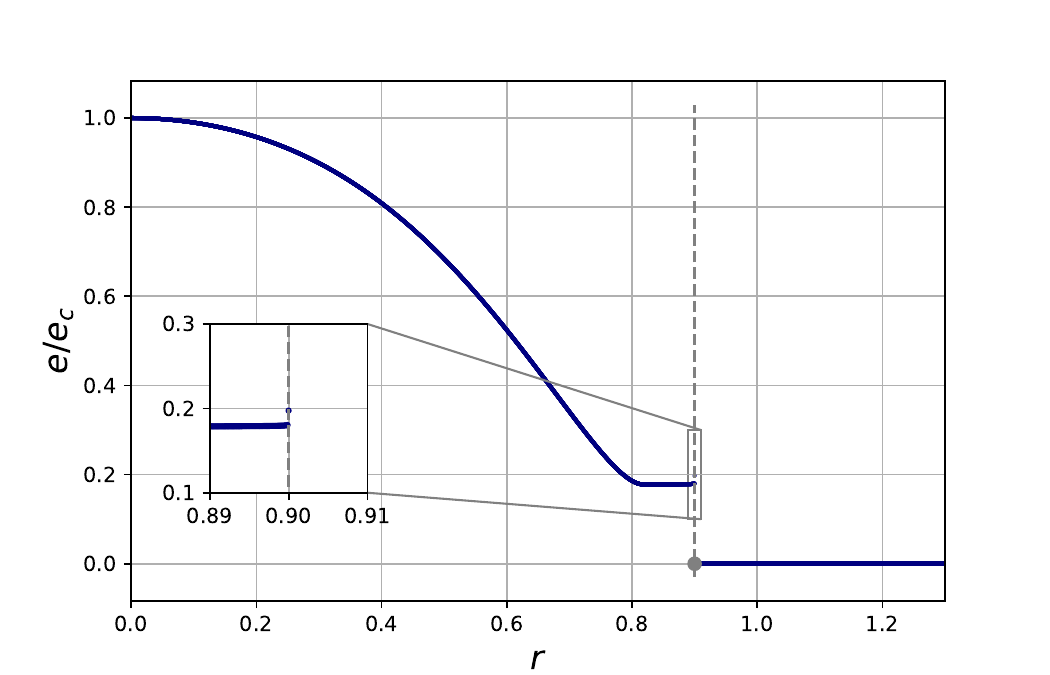}
         \caption{Specific internal energy for model 1.}
         \label{e0caso077}
     \end{subfigure}
     \hfill
     \begin{subfigure}[b]{0.49\textwidth}
         \centering
         \includegraphics[width=\textwidth]{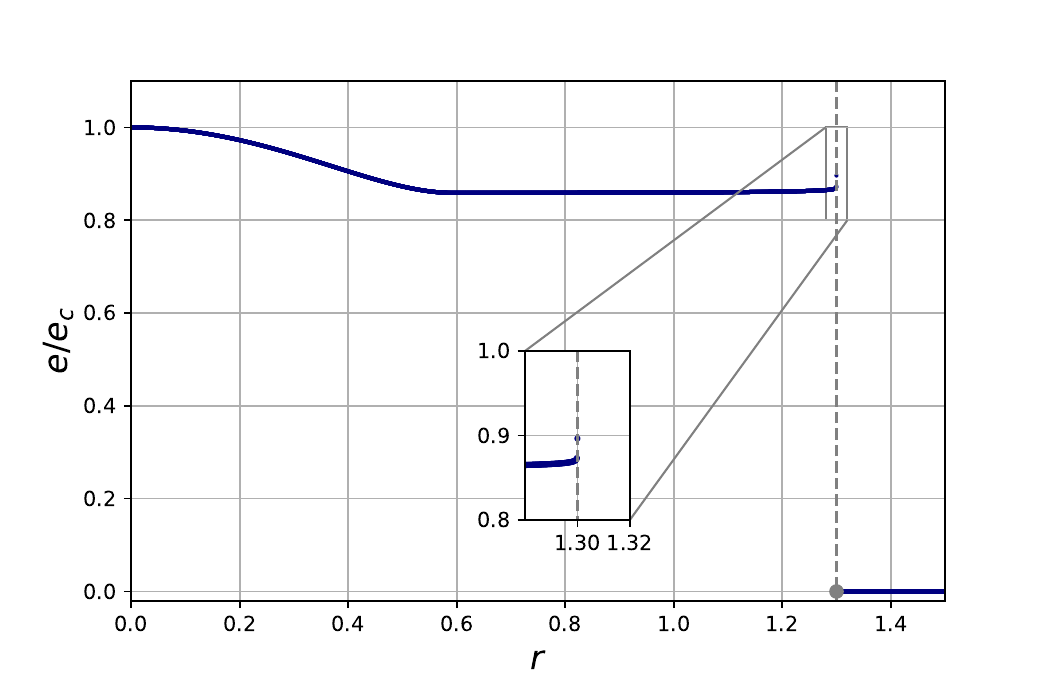}
         \caption{Specific internal energy for model 2.}
         \label{e0caso095}
     \end{subfigure}
        \caption{Specific internal energy profile $e(r)$ for the models presented in Table \ref{Cases}.}
        \label{e0}
\end{figure}

On the other hand, from Eq.~\eqref{TotRho} we can obtain the specific internal energy that is shown in Fig.~\ref{e0}.  As can be observed from the plots, there seems to be a small peak in the internal energy density as we approach the surface of the star (see inset in the plots). This peak is caused by numerical error, and it appears because as we approach the star's radius $R$, both the energy density and rest mass density approach zero, $\rho/\rho_0 \rightarrow 0/0$. However, $e(r)$ never diverges and the peak becomes smaller as we increase the numerical resolution. For $r > R$ we simply take $e=0$ outside the object since the specific internal energy density must vanish in the external \sch\ solution. 

\begin{figure}[h]
     \centering
     \begin{subfigure}[b]{0.49\textwidth}
         \centering
         \includegraphics[width=\textwidth]{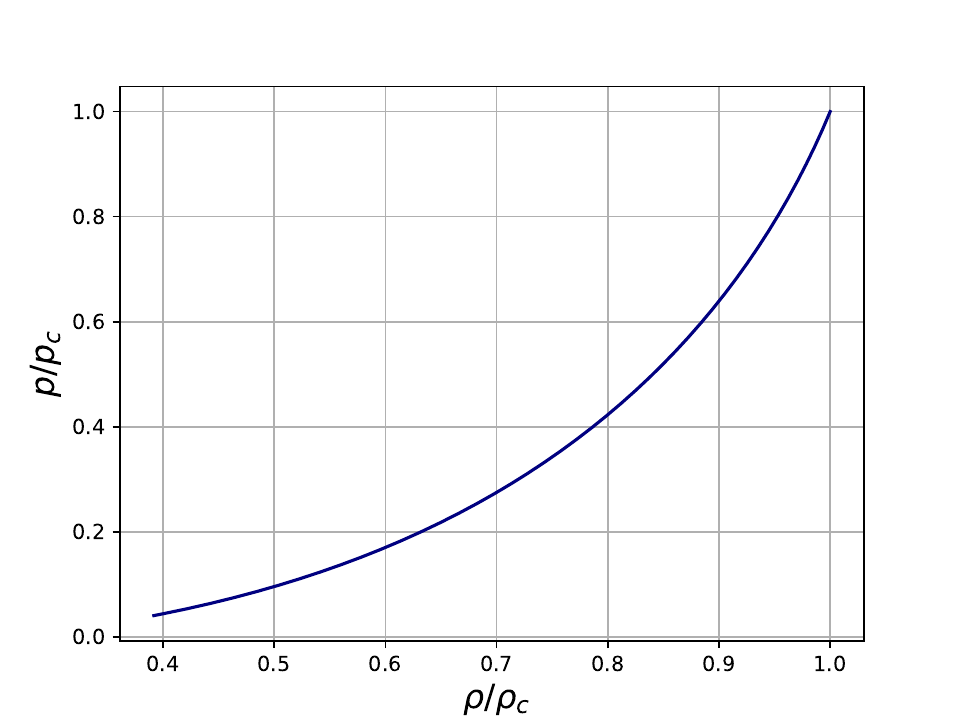}
         \caption{Pressure as a function of the total energy density for model 1.}
         \label{eos0caso077}
     \end{subfigure}
     \hfill
     \begin{subfigure}[b]{0.49\textwidth}
         \centering
         \includegraphics[width=\textwidth]{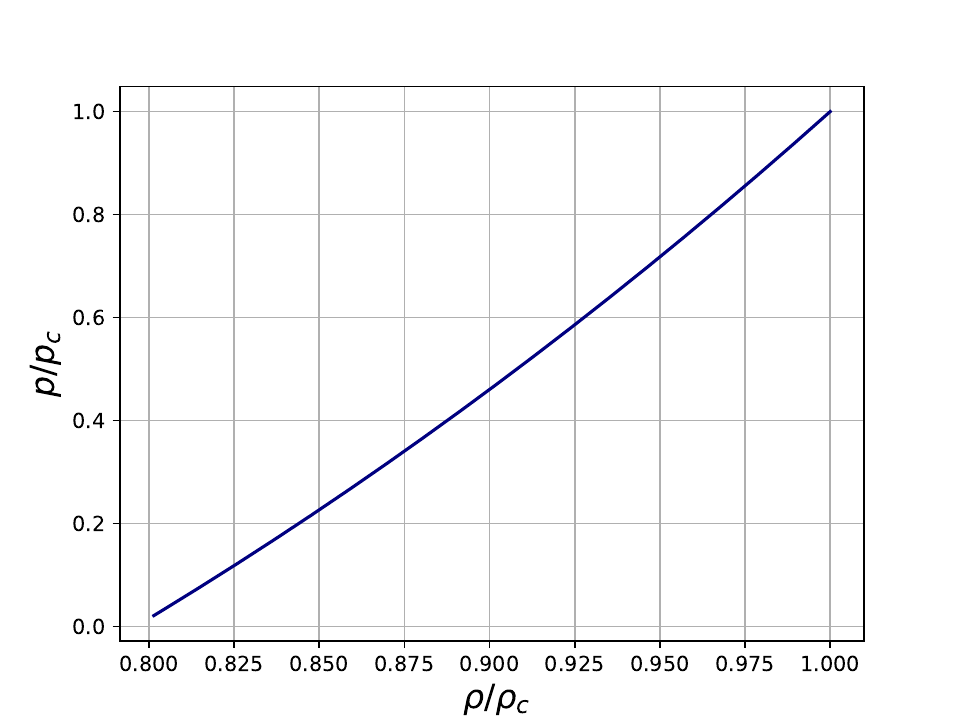}
         \caption{Pressure as a function of the total energy density for model 2.}
         \label{eos0caso095}
     \end{subfigure}
        \caption{Effective equation of state. Pressure $p$ as a function of the total energy density $\rho$ for the two models presented in Table~\ref{Cases}.}
        \label{eos0}
\end{figure}

Additionally, we can find an effective EoS by plotting the pressure as a function of the total energy density. The resulting graph is shown in Fig.~\ref{eos0}. This graph clearly illustrates the behavior at the core of the object. In the dust region, the EoS is simply $p=0$.

\begin{figure}[h]
     \centering
     \begin{subfigure}[b]{0.49\textwidth}
         \centering
         \includegraphics[width=\textwidth]{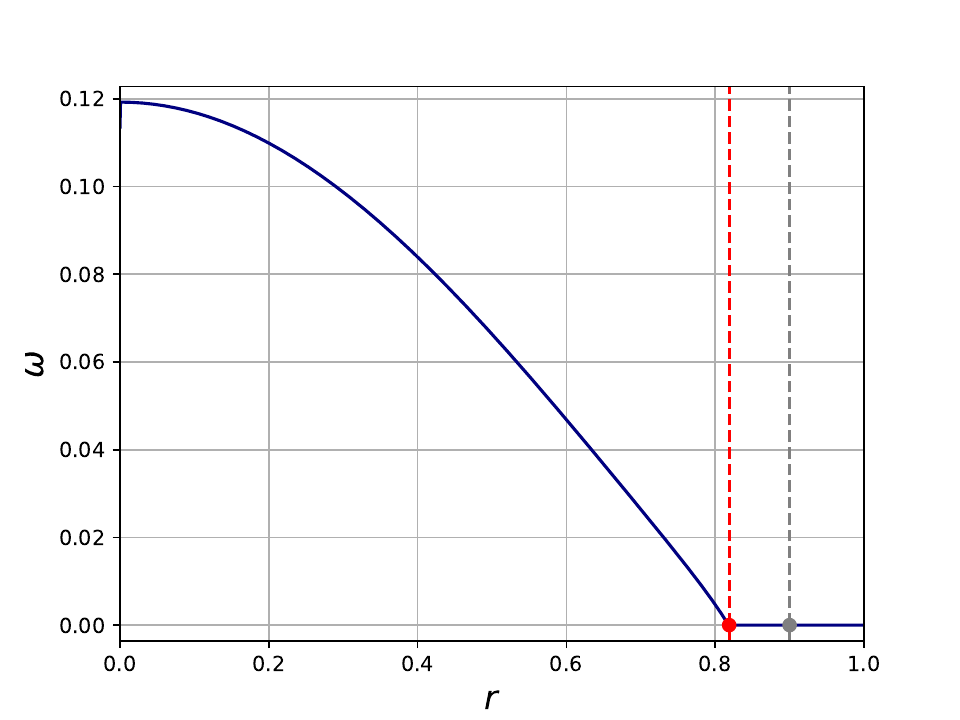}
         \caption{Baratropic function for model 1.}
         \label{omega0caso077}
     \end{subfigure}
     \hfill
     \begin{subfigure}[b]{0.49\textwidth}
         \centering
         \includegraphics[width=\textwidth]{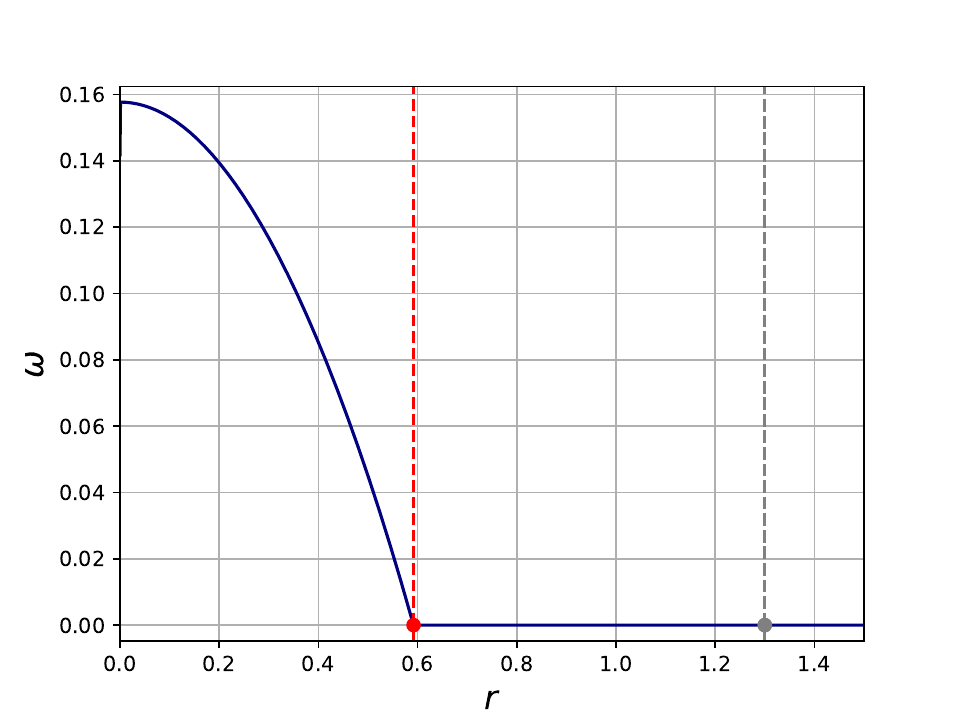}
         \caption{Baratropic function for model 2.}
         \label{omega0caso095}
     \end{subfigure}
        \caption{Ratio between the pressure and the total energy density $\omega(r)=p(r)/\rho(r)$ for the two models presented in Table \ref{Cases}.  The radius of the object is shown in gray.}
        \label{omega0}
\end{figure}

Figure~\ref{omega0} shows the barotropic function $\omega(r)=p(r)/\rho(r)$, which is the ratio between the pressure and the total energy density. The most common models used to study the interior solutions of spherically symmetric objects are based on barotropic or polytropic EoSs. In the barotropic case, the ratio between pressure and density remains constant, whereas in the polytropic case, the two quantities are related through a monotonic power-law relationship. In this study, we observe a behavior more akin to the polytropic case than to the barotropic case. The key point to emphasize here is that this behavior is derived from the numerical solution obtained from the field equations, and is no longer an assumption.

\begin{figure}[h]
     \centering
     \begin{subfigure}[b]{0.49\textwidth}
         \centering
         \includegraphics[width=\textwidth]{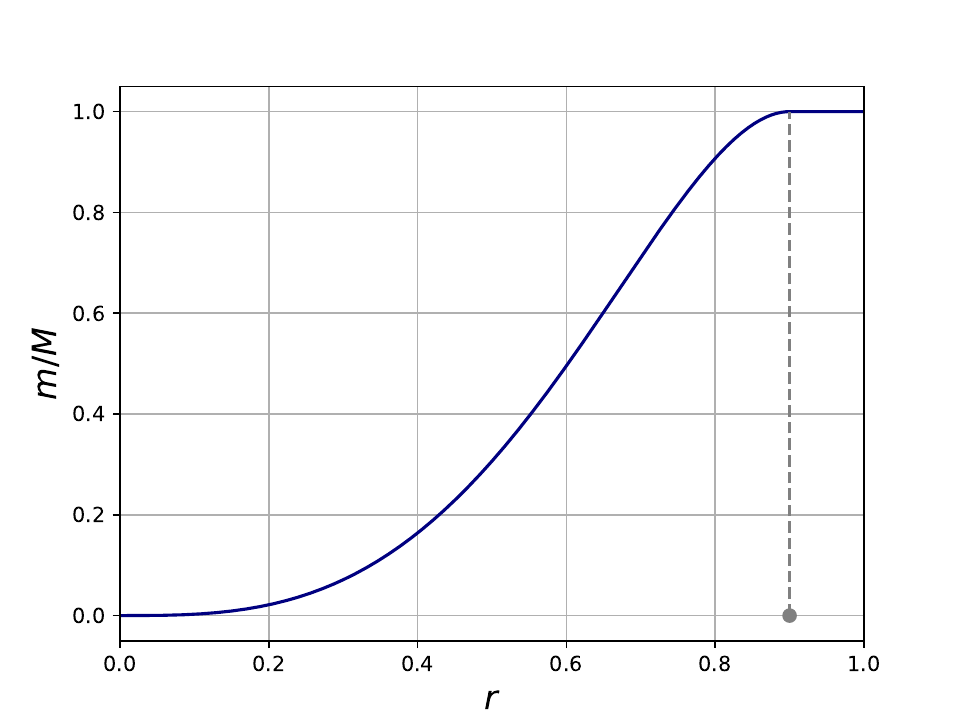}
         \caption{Mass function for model 1.}
         \label{mass0caso077}
     \end{subfigure}
     \hfill
     \begin{subfigure}[b]{0.49\textwidth}
         \centering
         \includegraphics[width=\textwidth]{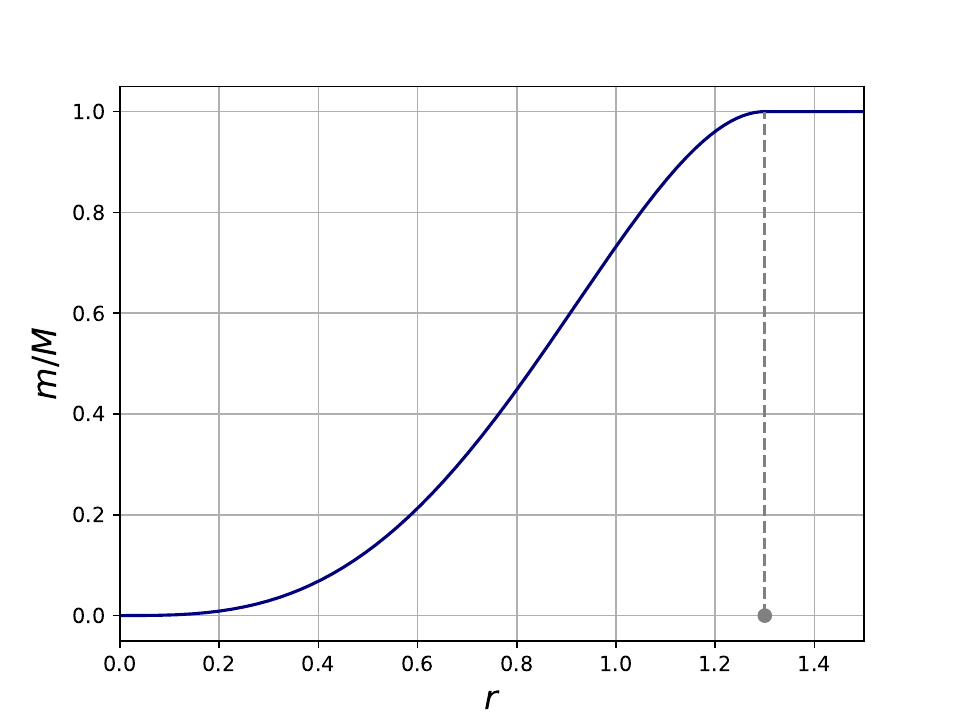}
         \caption{Mass function for model 2.}
         \label{mass0caso095}
     \end{subfigure}
        \caption{Mass function $m(r)$ for the two models presented in Table~\ref{Cases}. The radius of the object is shown in gray.}
        \label{mass0}
        
\end{figure}

\begin{figure}[h]
     \centering
     \begin{subfigure}[b]{0.49\textwidth}
         \centering
         \includegraphics[width=\textwidth]{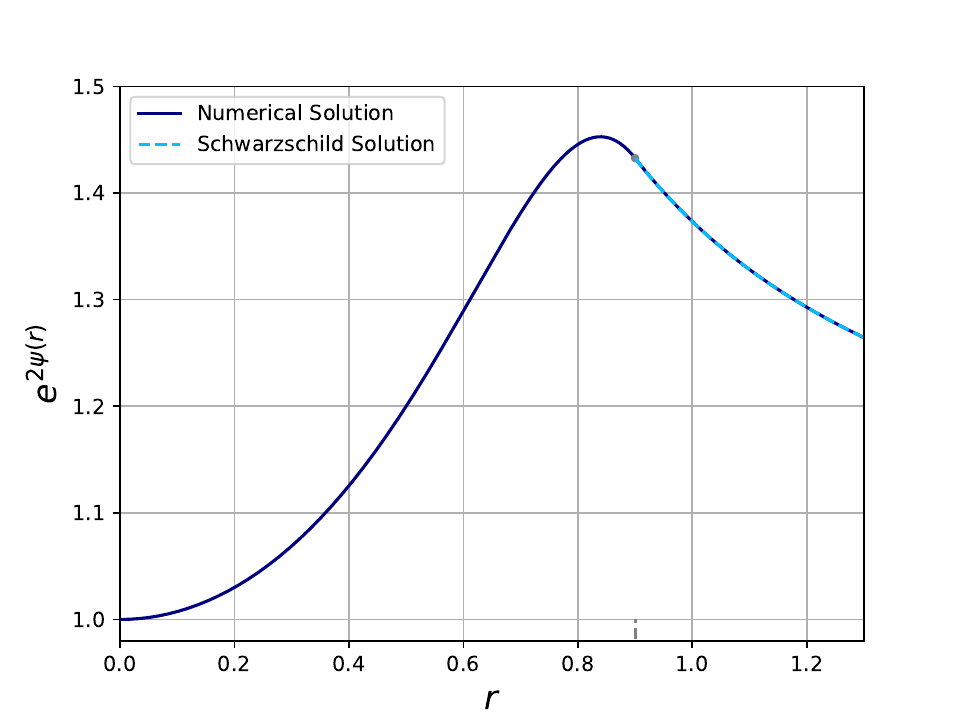}
         \caption{Metric function $A(r)$ for model 1.}
         \label{psi0caso077}
     \end{subfigure}
     \hfill
     \begin{subfigure}[b]{0.49\textwidth}
         \centering
         \includegraphics[width=\textwidth]{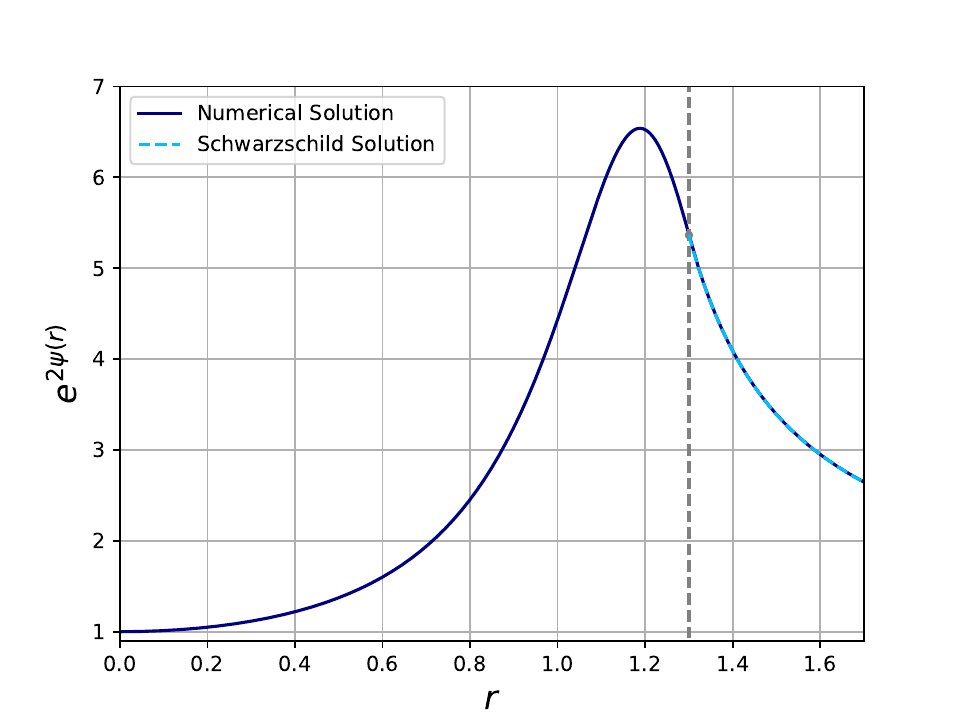}
         \caption{Metric function $A(r)$ for model 2.}
         \label{psi0caso095}
     \end{subfigure}
        \caption{Radial metric function $A(r)=e^{2\psi(r)}$ for the two models presented in Table~\ref{Cases}. The radius of the object is shown in gray.}
        \label{psi}
\end{figure}

\begin{figure}[h]
     \centering
     \begin{subfigure}[b]{0.49\textwidth}
         \centering
         \includegraphics[width=\textwidth]{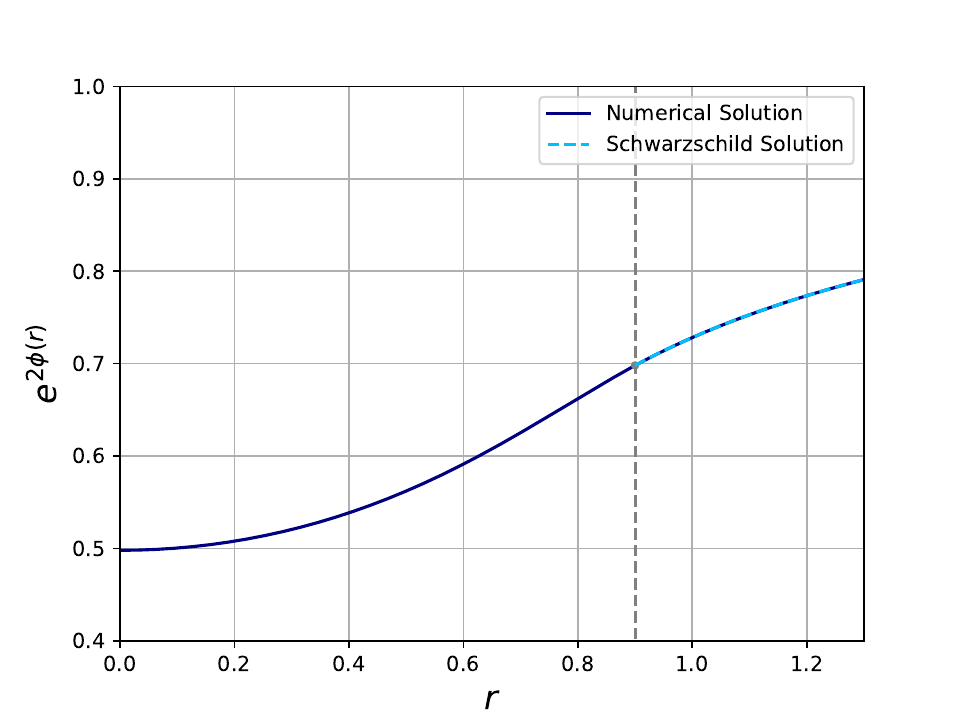}
         \caption{Metric function $e^{\phi(r)}$ for model 1.}
         \label{PHIcaso077}
     \end{subfigure}
     \hfill
     \begin{subfigure}[b]{0.49\textwidth}
         \centering
         \includegraphics[width=\textwidth]{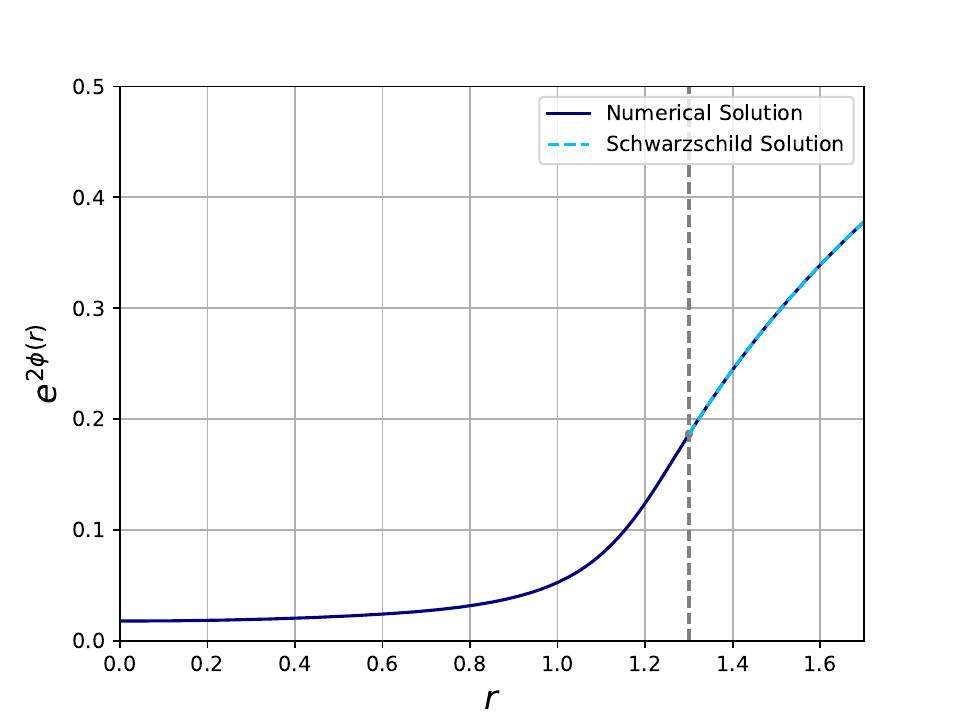}
         \caption{Metric function $e^{2\phi(r)}$ for model 2.}
         \label{PHIcaso095}
     \end{subfigure}
        \caption{Metric function $e^{2\phi(r)}$ for the two models presented in Table~\ref{Cases}. The radius of the object is shown in gray.}
        \label{phi}
\end{figure}

Finally, the mass function $m(r)$ is presented in Fig.~\ref{mass0}, and the metric functions $e^{2\phi(r)}$ and $A(r)=e^{2\psi(r)}$ in Figs.~\eqref{psi} and \eqref{phi}, respectively. In both cases, the Buchdahl condition~\eqref{acond} is clearly satisfied, as evidenced by the fact that the maximum value of the metric function $A(r)$ is less than $9$. The metric function $g_{00}=e^{2\phi(r)}$ is also interesting, as it represents the gravitational time dilation for a stationary observer, $d \tau = e^{\phi(r)} dt$, and it also plays the role of the Newtonian gravitational potential.  Notice that for model 1 we already have $e^{2\phi(r=0)} \simeq 0.5$ at the center of the star, while for model 2 we have $e^{2\phi(r=0)} \simeq 0.02$ indicating an extremely large gravitational time dilation factor. Furthermore, it should be noted that the metric functions match the well-known Schwarzschild solution outside the star's radius (shown as a dotted blue line in the plots).

\section{Case of canonical neutron stars}
\label{sec:NS}

In this section, we will see if our model of a spherically symmetric mass distribution can be applied to the case of realistic neutron stars. Consider first a sphere of uniform density with mass $M = 1.4 \: \mathrm{M}_\odot$, radius $R = 12 \: \mathrm{km}$, and volumetric mean density $\bar{\rho}_0=10^{17} \: \mathrm{kg/m}^3$, which is the density of an atomic nucleus. The above parameters are used to characterize canonical neutron stars (see for example~\cite{condon2016essential}). Moreover, we assume that the central pressure is also determined by the pressure of an atomic nucleus, which is around $5.24 \times 10^{33} \: \mathrm{N/m}^2$. 

By using the above information, it is possible to fix three of our five free parameters. First, the radius of the star is set to $R=12 \: \mathrm{km}$. Second, in order to find the rest energy density we use the volumetric density in the following way: $\rho_{0_c} = c^2\bar{\rho_0} = 3.48 \times 10^{34} \: \mathrm{J/m}^3$, which corresponds to $\rho_{0_c} = 2.87 \times 10^{-4} \: \mathrm{km}^{-2}$ in geometric units.  Finally, the pressure $p_c = 5.24\times 10^{33} \: \mathrm{N/m}^2$ becomes $p_c =  4.33\times10^{-5} \: \mathrm{km}^{-2}$ in geometric units.

\begin{figure}[h]
     \centering
     \begin{subfigure}[b]{0.49\textwidth}
         \centering
         \includegraphics[width=\textwidth]{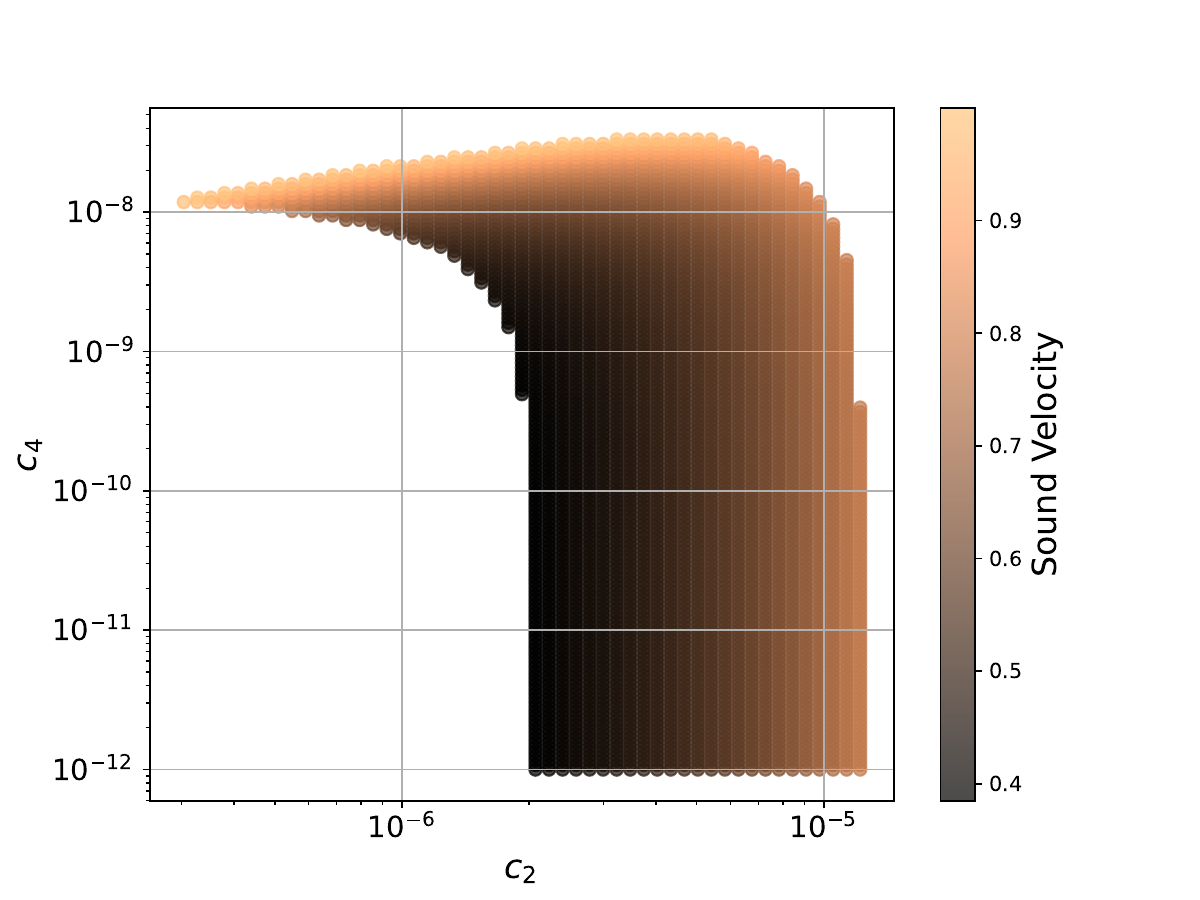}
         \caption{Sound speed for the allowed cases.}
         \label{SoundVel}
     \end{subfigure}
     \hfill
     \begin{subfigure}[b]{0.49\textwidth}
         \centering
         \includegraphics[width=\textwidth]{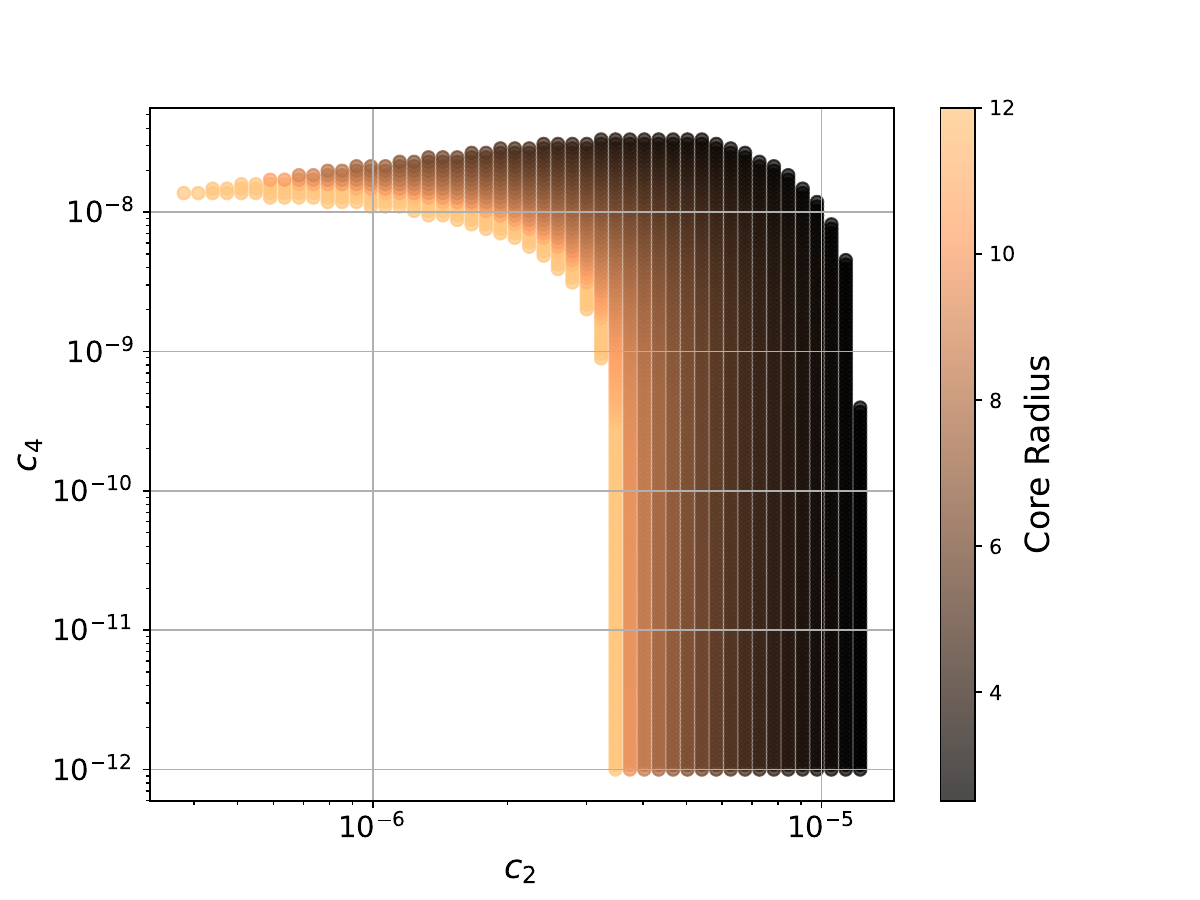}
         \caption{Core radius for the allowed cases.}
         \label{CoreRadius}
     \end{subfigure}
        \caption{Values of the parameters $c_2$ and $c_4$ that meet the physical requirements established in subsection~\ref{sec:PhysicalCond}. The color map of the subplot (\ref{SoundVel}) shows the sound speed at the center of the object, while the color map of the subplot (\ref{CoreRadius}) shows the size of the core radius of the numerical solution.}
        \label{canNS}
\end{figure}

With this set of fixed parameters we can now find the interval of values for $c_2$ and $c_4$ that are consistent with the physical requirements of sections \ref{subsec:r0} and \ref{sec:PhysicalCond}. As a result, we obtain the regions presented in Fig.~\ref{canNS}. The information about the sound speed is shown in Figure~\ref{SoundVel}. Additionally, Figure~\ref{CoreRadius} presents the thickness of the object's central core where the pressure is non-zero, i.e. the "core radius". Specifically, we have chosen to represent the thickness of the central core ignoring the dust layer. In particular, we can observe that the core radius has larger values when the sound speed at the center is smaller.

\begin{figure}[h]
     \centering
     \begin{subfigure}[b]{0.49\textwidth}
         \centering
         \includegraphics[width=\textwidth]{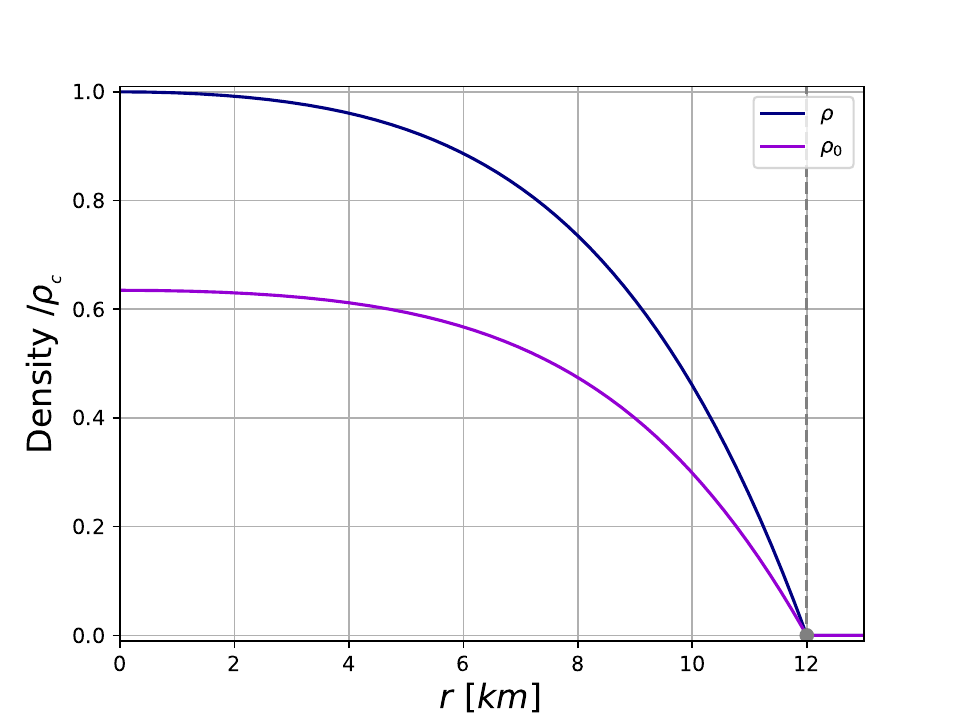}
         \caption{Energy densities for model 3.}
         \label{rho0caso1NS}
     \end{subfigure}
     \hfill
     \begin{subfigure}[b]{0.49\textwidth}
         \centering
         \includegraphics[width=\textwidth]{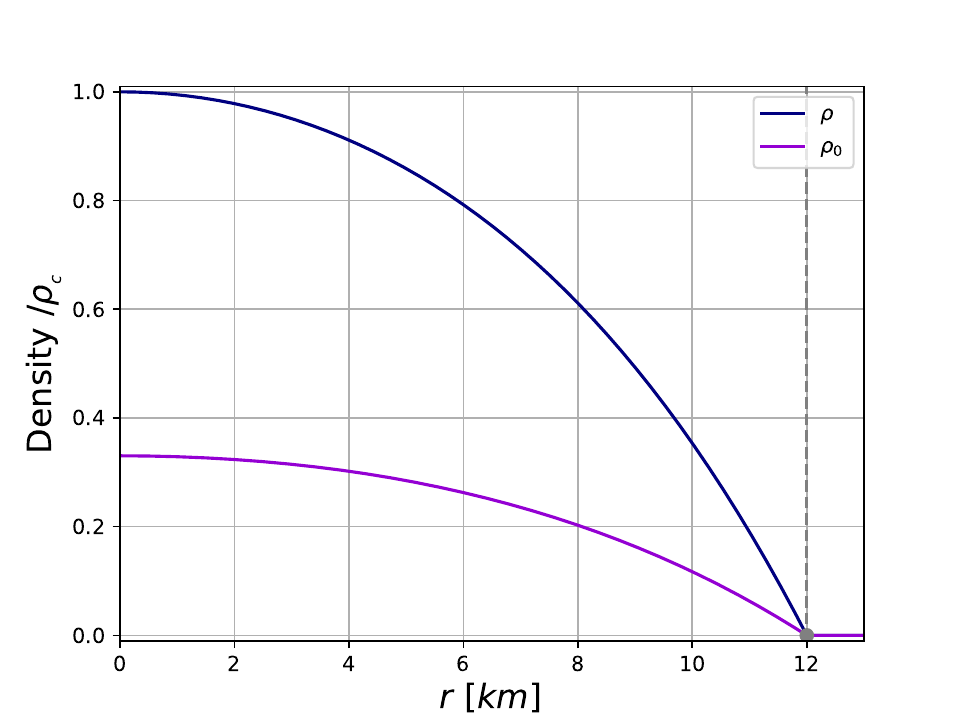}
         \caption{Energy densities for model 4.}
         \label{rho0caso2NS}
     \end{subfigure}
\caption{Total and rest mass energy densities profiles $\rho(r)$ and $\rho_0$ for the two models presented in Table~\ref{CasesNS}. The radius of the star is represented by the vertical gray line.}
\label{rho_NS}
\end{figure}

We will now consider the two specific models presented in Table~\ref{CasesNS}. The numerical solution of the equations is represented in the following set of plots, where we can see that they have a behavior similar to the models presented in the previous section. Figure~\ref{rho_NS} illustrates the behavior of the total energy density and the rest mass energy density for the parameters given in Table \ref{CasesNS}. We observe a smooth and monotonically decreasing behavior for the densities, which vanish at the radius of the star as required.

\begin{table}
    \centering
    \begin{tabular}{|l|c|c|}
    \hline
    \textbf{Parameter$\ \ \ \ \ $} & \textbf{Model 3}&\textbf{Model 4} \\
        \hline
        \hline
          Constant $c_2$  & $8.55\times 10^{-7}$ &$4.67\times 10^{-6}$\\
          Constant $c_4$  & $1.59\times 10^{-8}$ &$9.46\times 10^{-9}$\\
         \hline
    \end{tabular}
    \centering
\caption{Two sets of parameters necessary for integrating the TOV equations with radius $R=12 \: \mathrm{km}$. Model 3 leads to a central sound speed $v_s|_{r=0} =0.84$, while for model 4 we obtain $v_s|_{r=0} =0.64$. In both cases the central rest energy density is $\rho_{0_c} = 2.87 \times 10^{-4} \: \mathrm{km}^{-2}$ and the central pressure is $p_c = 4.33\times 10^{-5} \: \mathrm{km}^{-2}$.}
\label{CasesNS}
\end{table}

\begin{figure}[h]
     \centering
     \begin{subfigure}[b]{0.49\textwidth}
         \centering
         \includegraphics[width=\textwidth]{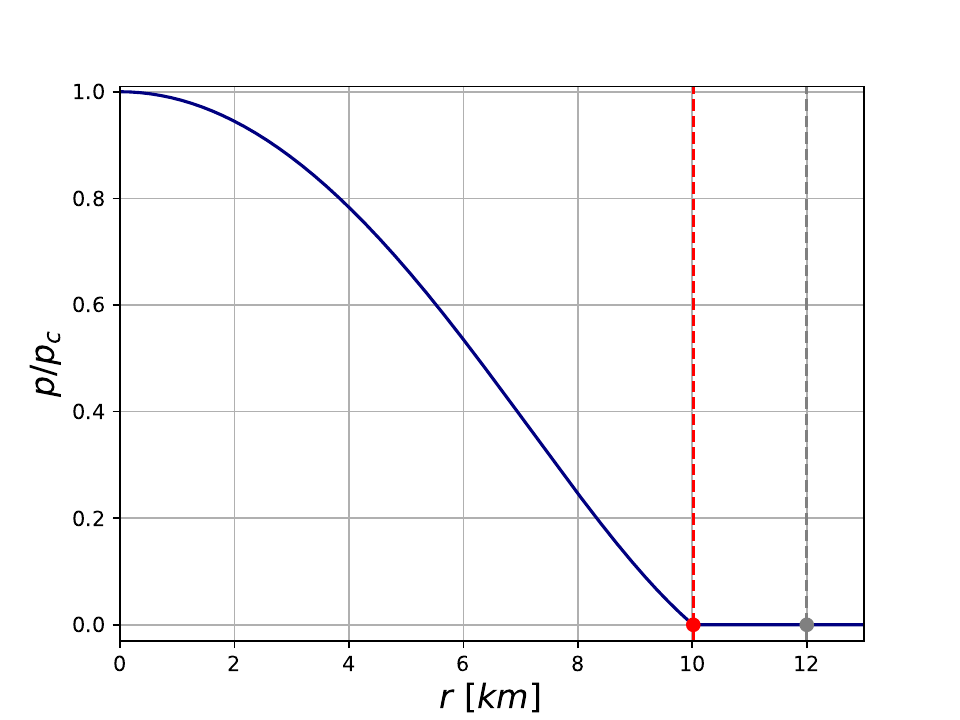}
         \caption{Pressure for model 3. The radius or zero-pressure is $r_{p1}=10.02$.}
         \label{p0caso1NS}
     \end{subfigure}
     \hfill
     \begin{subfigure}[b]{0.49\textwidth}
         \centering
         \includegraphics[width=\textwidth]{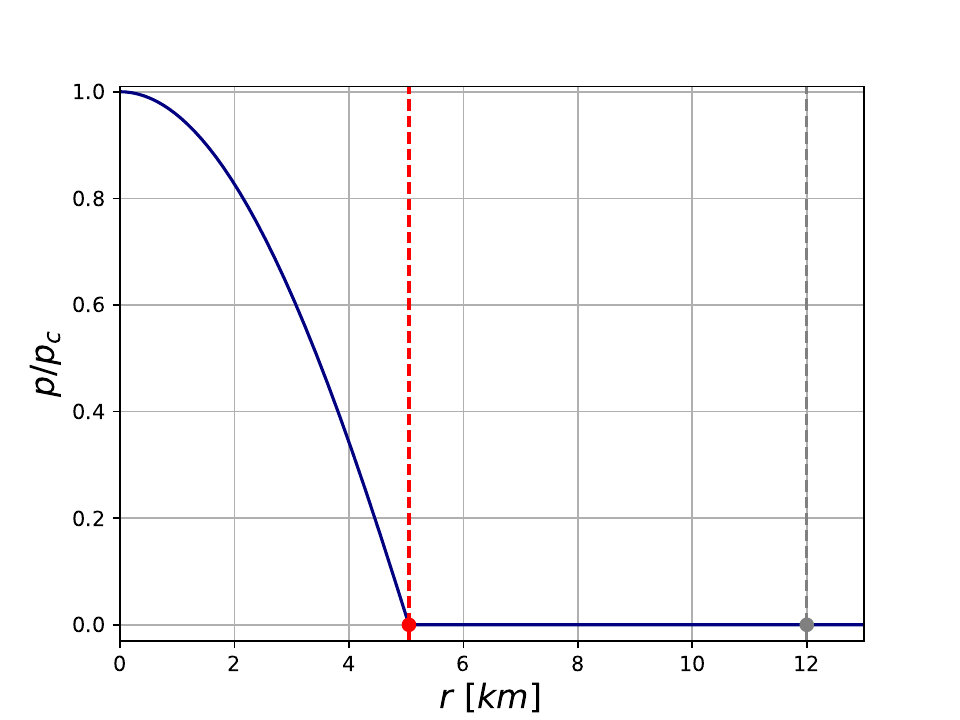}
         \caption{Pressure for model 4. The radius of 
         zero-pressure is  $r_{p2}=5.05\ km$.}
         \label{p0caso2NS}
     \end{subfigure}
\caption{Pressure profile $p(r)$ for the two models presented in Table~\ref{CasesNS}. The radius of the star corresponds to the vertical gray line, and the radius where the pressure vanishes corresponds to the vertical red line.}
\label{p0NS}
\end{figure}

The dependency of the pressure $p$ on the radial coordinate $r$ is shown in Fig.~\ref{p0NS}. The pressure vanishes at $r_{p1} = 10.02 \: \mathrm{km}$ for model 3, and at $r_{p2} = 5.05 \: \mathrm{km}$ for model 4. Thus, in both cases, the central core is surrounded by a layer of dust where the pressure vanishes. The layer of dust has a thickness of $1.98 \: \mathrm{km}$ and $6.95 \: \mathrm{km}$ in models 3 and 4, respectively.

\begin{figure}[h]
     \centering
     \begin{subfigure}[b]{0.49\textwidth}
         \centering
         \includegraphics[width=\textwidth]{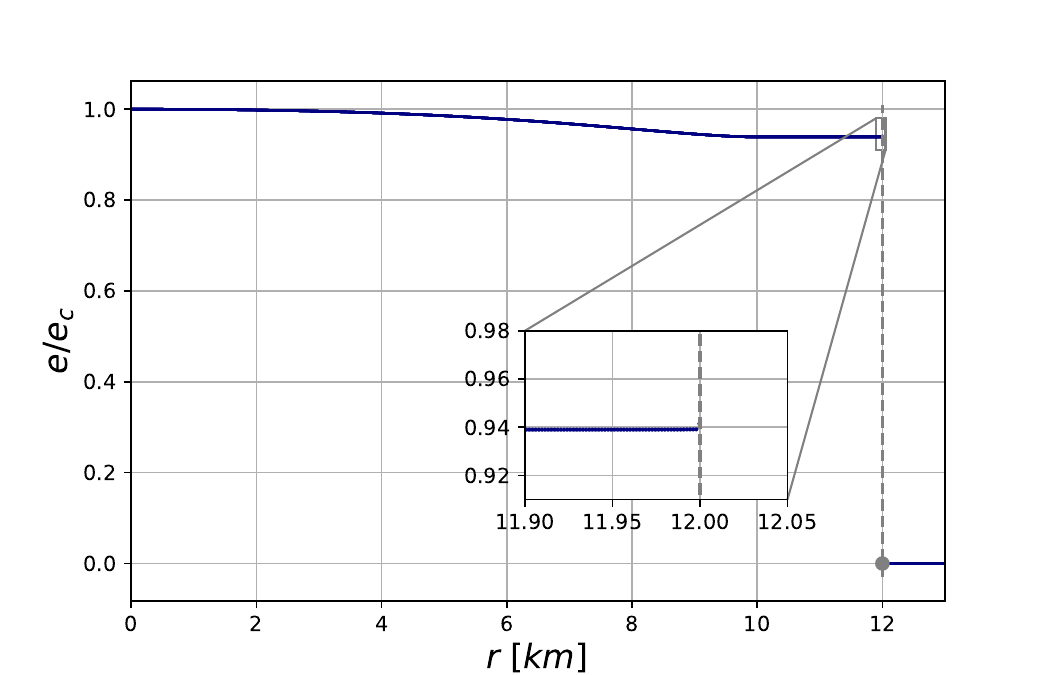}
         \caption{Specific internal energy for model 3.}
         \label{e0NS1}
     \end{subfigure}
     \hfill
     \begin{subfigure}[b]{0.49\textwidth}
         \centering
         \includegraphics[width=\textwidth]{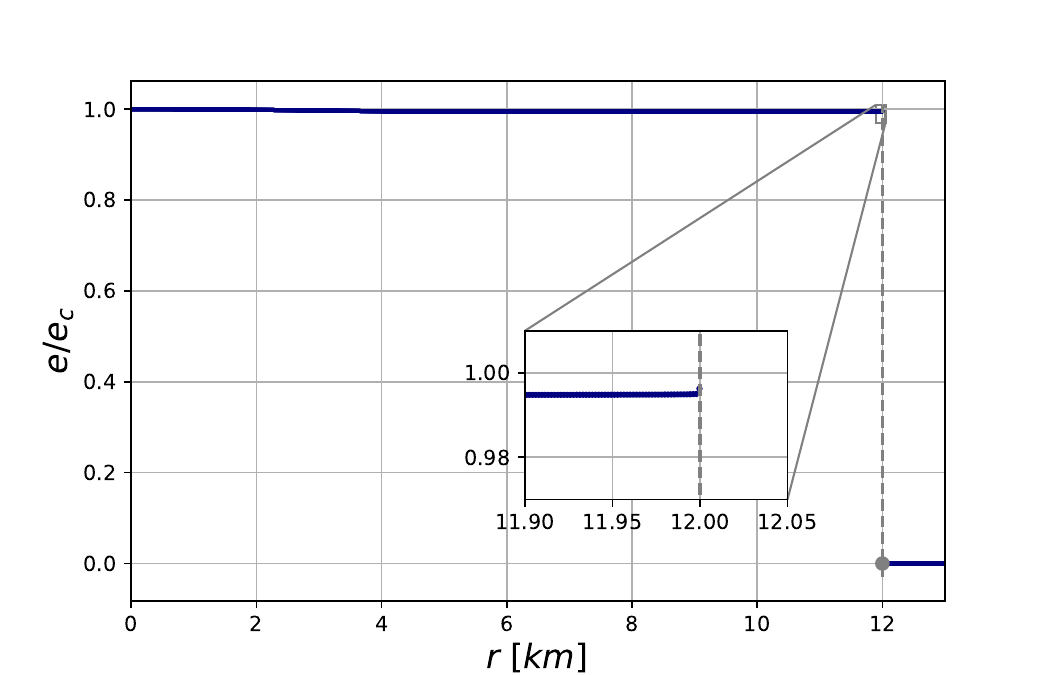}
         \caption{Specific internal energy for model 4.}
         \label{e0NS2}
     \end{subfigure}
\caption{Specific internal energy profile $e(r)$ for the models presented in Table~\ref{CasesNS}. The radius of the star is shown as a gray vertical dashed line.}
\label{e0NS}
\end{figure}

The specific internal energy profile $e(r)$ is presented in Fig.~\ref{e0NS}. As before, as we approach the star's radius $R$ there is a small peak in $e(r)$ caused by a numerical error coming from the fact that both $\rho$, and $\rho_0$ vanish there,  but the internal energy remains finite.

\begin{figure}[h]
     \centering
     \begin{subfigure}[b]{0.49\textwidth}
         \centering
         \includegraphics[width=\textwidth]{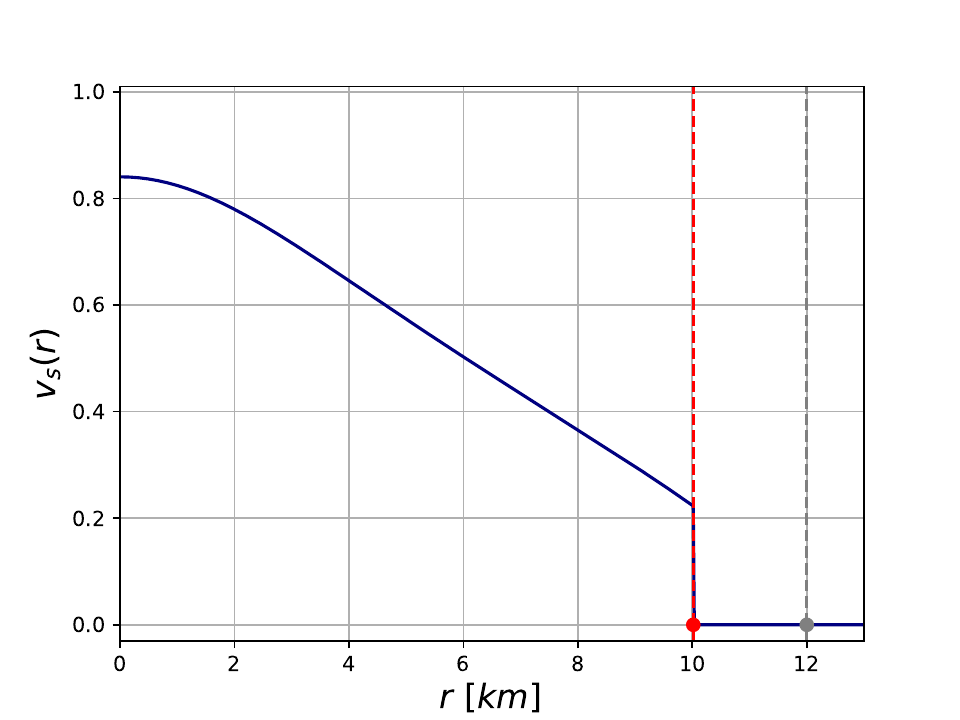}
         \caption{Sound speed in model 3.}
         \label{vs0caso1NS}
     \end{subfigure}
     \hfill
     \begin{subfigure}[b]{0.49\textwidth}
         \centering
         \includegraphics[width=\textwidth]{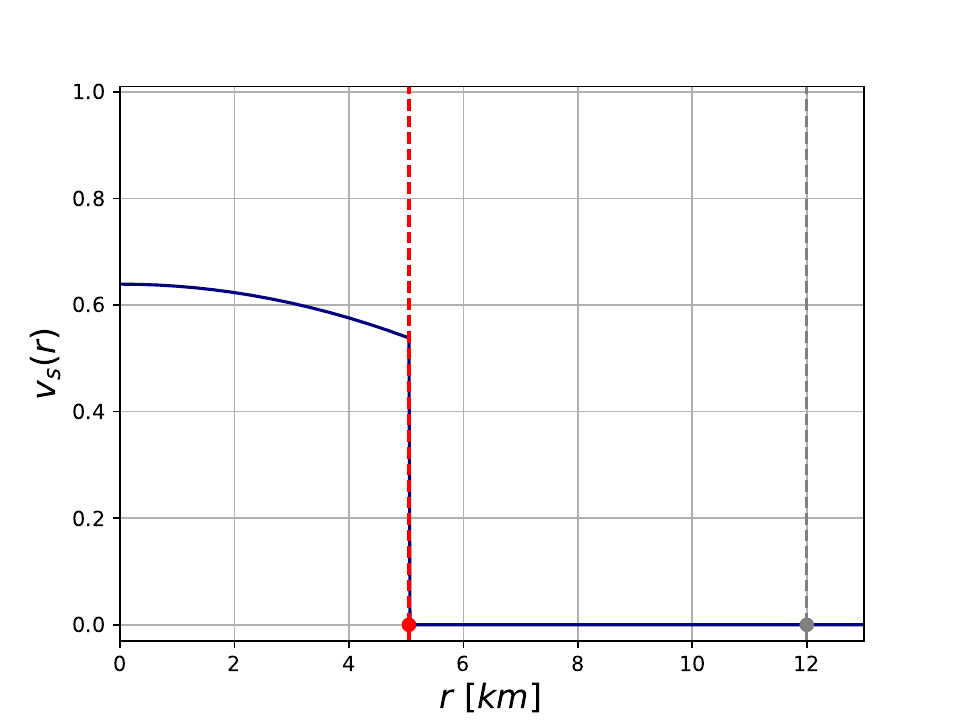}
         \caption{Sound speed in model 4.}
         \label{vs0caso2NS}
     \end{subfigure}
\caption{Sound speed inside the star $v_s(r)$ for the two models presented in Table~\ref{CasesNS}. The red and gray vertical dashed lines represent the radius of zero pressure and the radius of the star, respectively.}
\label{vs0NS}
\end{figure}

The speed of sound for both models is shown in Fig.~\ref{vs0NS}.  Its maximum value is attained at the center of the star, with $v_s=0.84$ for model 3 and $v_s=0.64$ for model 4. In both models, the speed of sound decreases monotonically and the speed of light limit is fully satisfied, as expected.

\begin{figure}[h]
     \centering
     \begin{subfigure}[b]{0.49\textwidth}
         \centering
         \includegraphics[width=\textwidth]{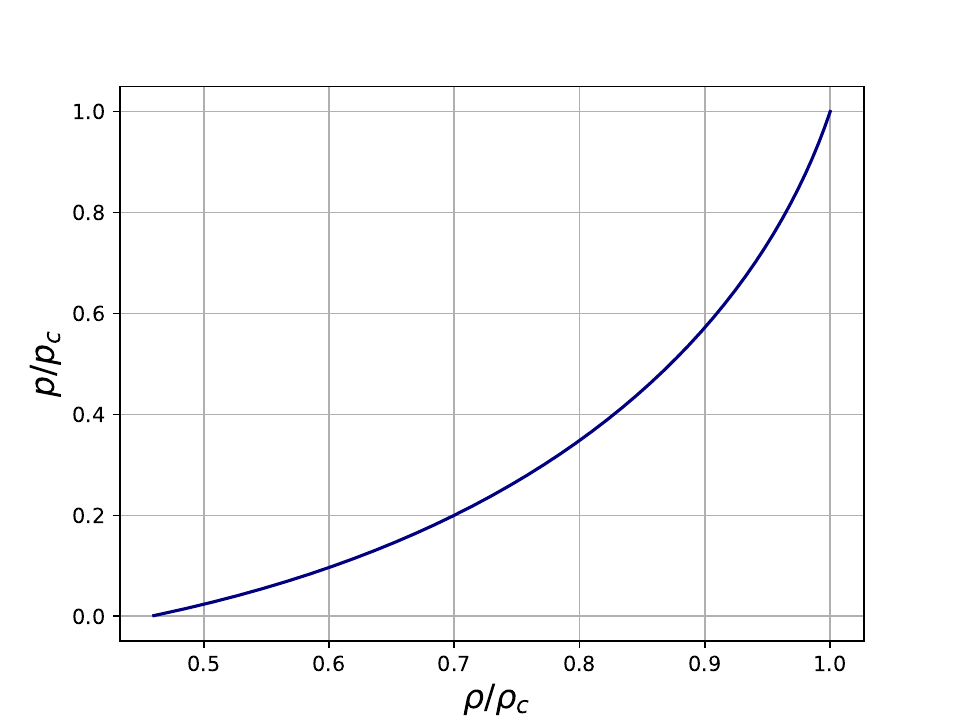}
         \caption{Equation of state for model 3.}
         \label{eos0caso1NS}
     \end{subfigure}
     \hfill
     \begin{subfigure}[b]{0.49\textwidth}
         \centering
         \includegraphics[width=\textwidth]{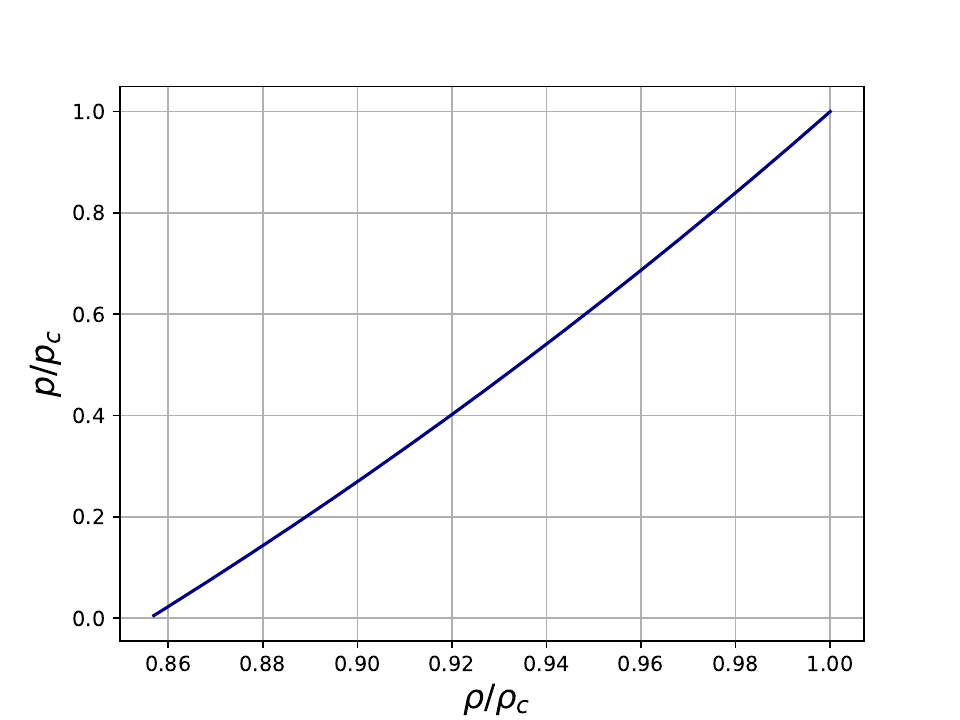}
         \caption{Equation of state for model 4.}
         \label{eos0caso2NS}
     \end{subfigure}
\caption{Equation of State. Pressure $p$ as a function of the total energy density $\rho$ for the two models presented in Table~\ref{CasesNS}.}
\label{eos0NS}
\end{figure}

As before, we can derive an EoS by plotting the pressure as a function of the total energy density for the core. This is shown in Fig.~\ref{eos0NS}, where we observe that the pressure is a monotonically increasing function of the density.

\begin{figure}[h]
     \centering
     \begin{subfigure}[b]{0.49\textwidth}
         \centering
         \includegraphics[width=\textwidth]{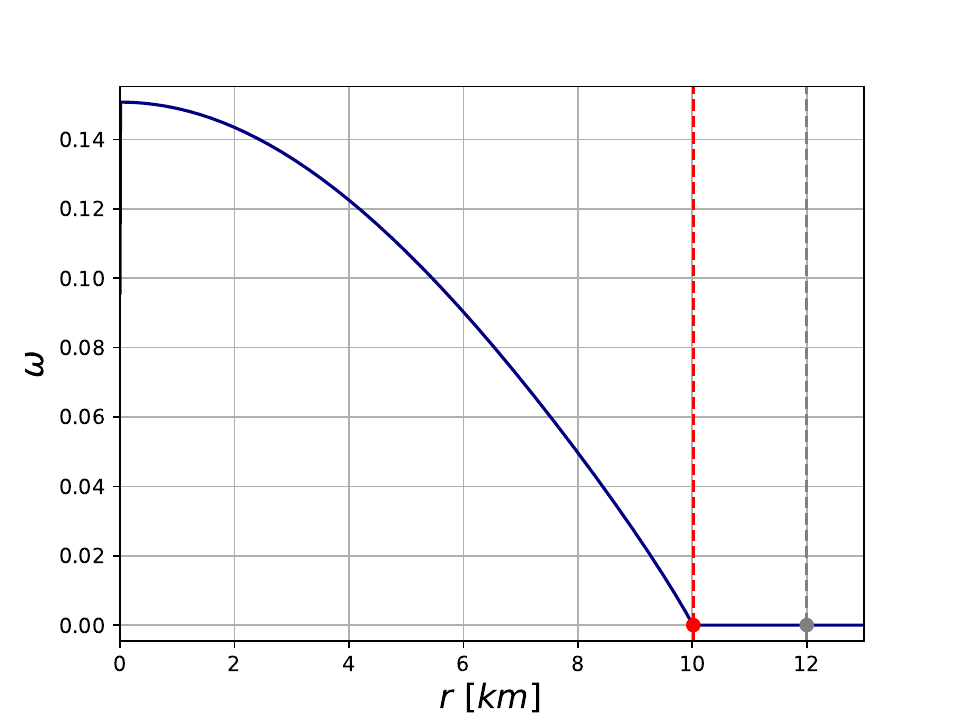}
         \caption{Ratio of pressure to density for model 3.}
         \label{omega01NS}
     \end{subfigure}
     \hfill
     \begin{subfigure}[b]{0.49\textwidth}
         \centering
         \includegraphics[width=\textwidth]{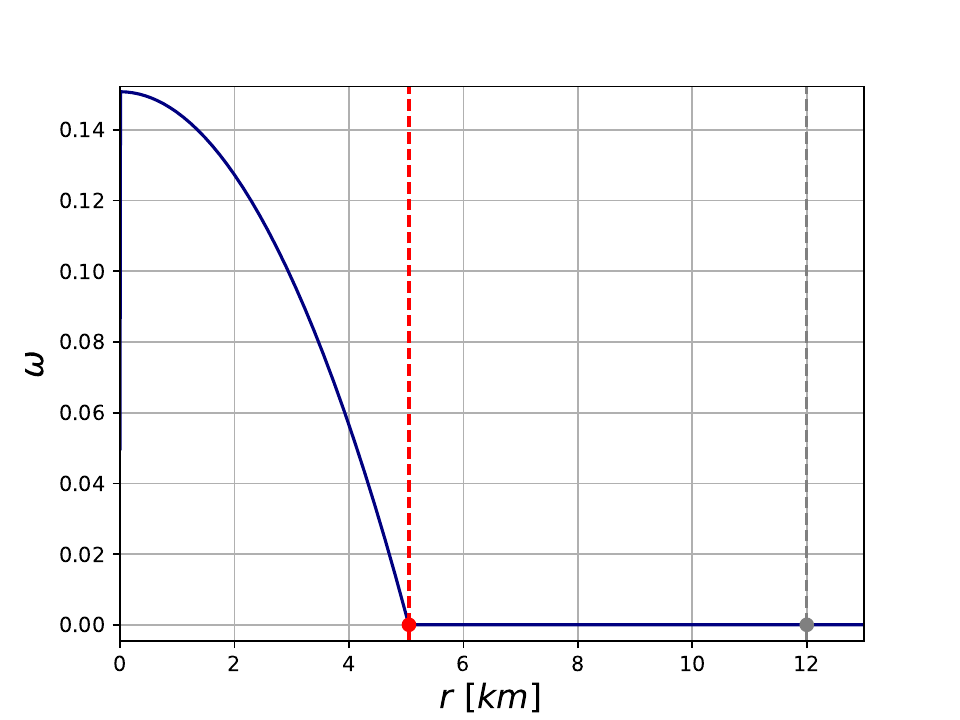}
         \caption{Ratio of pressure to density for model 4.}
         \label{omega02NS}
     \end{subfigure}
\caption{Ratio between the pressure and the total energy density, $\omega(r)=P(r)/\rho(r)$ ,for the two models presented in Table~\ref{CasesNS}. The radius of the star is shown as a vertical gray line.}
\label{omega0NS}
\end{figure}

Figure~\ref{omega0NS} shows the ratio $\omega(r)=p(r)/\rho(r)$, which is again a result of the numerical solution and exhibits behavior similar to the polytropic case. Additionally, the mass function profile $m(r)$, shown in Fig.~\ref{mass0}, becomes constant for radii greater than the star's radius, i.e., in the exterior of the spherical object.

\begin{figure}[h]
     \centering
     \begin{subfigure}[b]{0.49\textwidth}
         \centering
         \includegraphics[width=\textwidth]{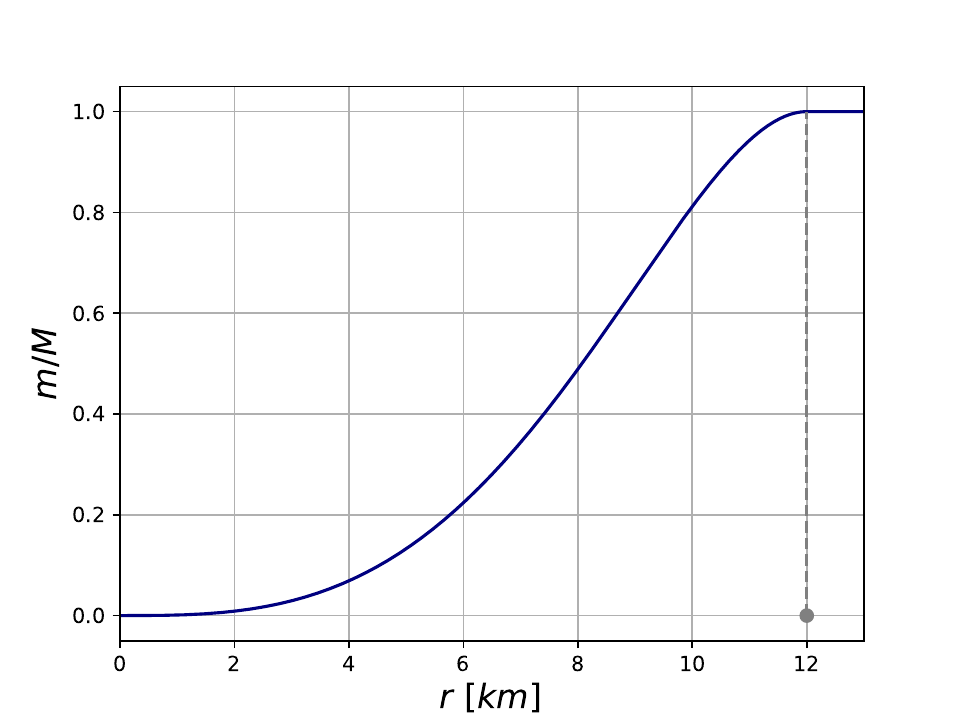}
         \caption{Mass function for model 3.}
         \label{mass01NS}
     \end{subfigure}
     \hfill
     \begin{subfigure}[b]{0.49\textwidth}
         \centering
         \includegraphics[width=\textwidth]{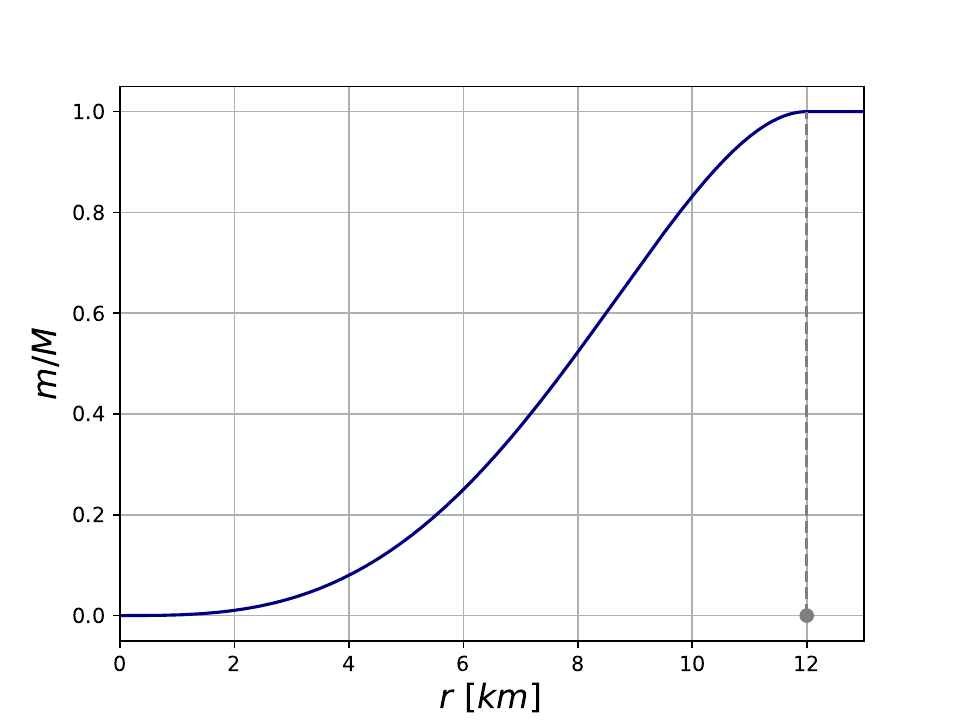}
         \caption{Mass function for model 4.}
         \label{mass02NS}
     \end{subfigure}
\caption{Mass function $m(r)$ for the two models presented in Table~\ref{CasesNS}. The radius of the star is denoted by a vertical gray line.}
\label{mass0NS}
\end{figure}

\begin{figure}[h]
     \centering
     \begin{subfigure}[b]{0.49\textwidth}
         \centering
         \includegraphics[width=\textwidth]{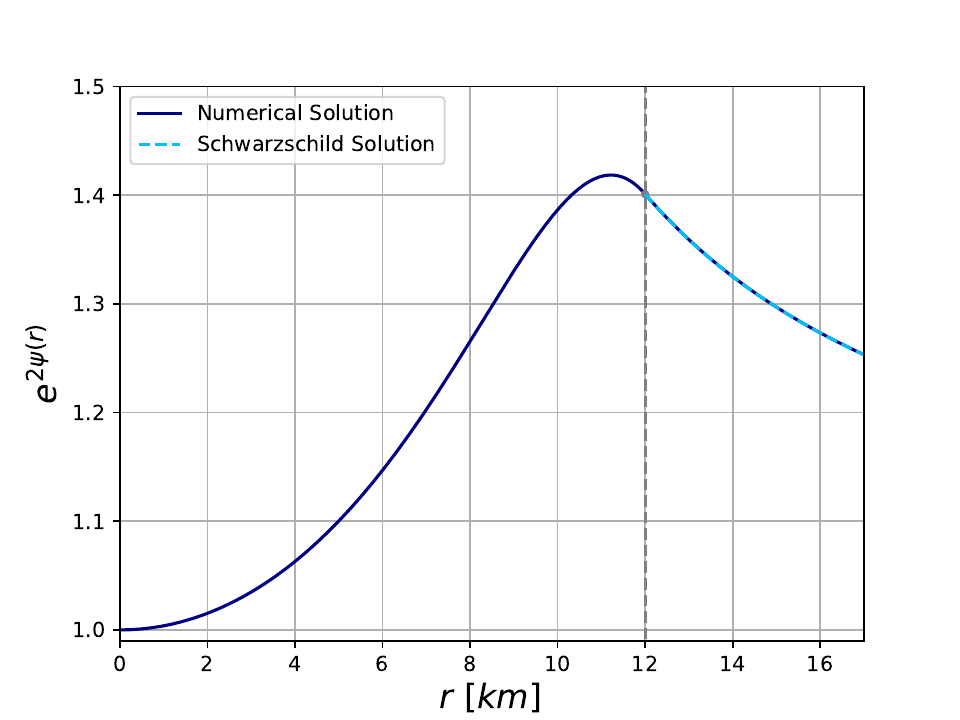}
         \caption{Metric function $A(r)=e^{2\psi(r)}$ for model 3.}
         \label{psi0caso077NS}
     \end{subfigure}
     \hfill
     \begin{subfigure}[b]{0.49\textwidth}
         \centering
         \includegraphics[width=\textwidth]{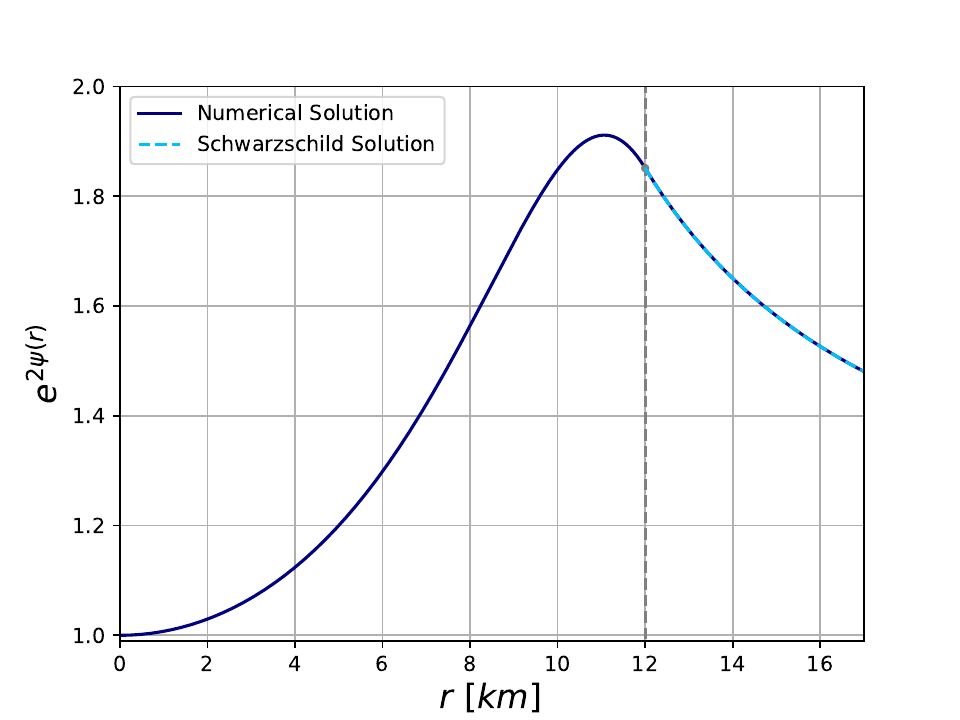}
         \caption{Metric function $A(r)=e^{2\psi(r)}$ for model 4.}
         \label{psi0caso095NS}
     \end{subfigure}
\caption{Metric function $A(r)=e^{2\psi(r)}$ for the two models presented in Table~\ref{CasesNS}. The radius of the object is shown in gray.}
\label{psi0NS}
\end{figure}

\begin{figure}[h]
     \centering
\begin{subfigure}[b]{0.49\textwidth}
         \centering
         \includegraphics[width=\textwidth]{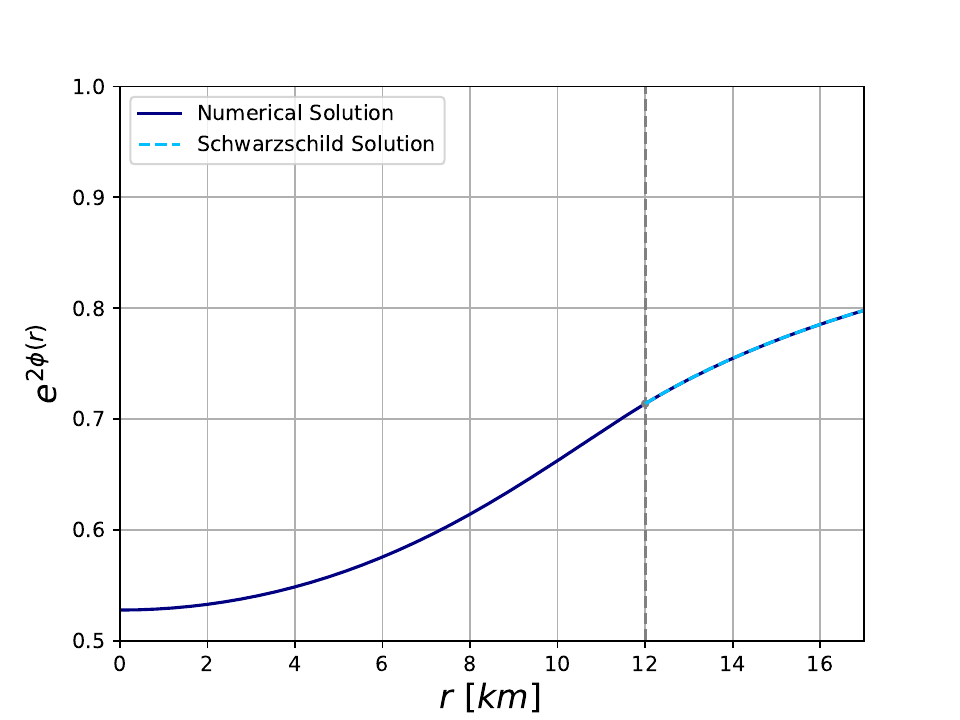}
         \caption{Metric function $e^{2\phi(r)}$ in model 3.}
         \label{phi_c1}
     \end{subfigure}
     \hfill
     \begin{subfigure}[b]{0.49\textwidth}
         \centering
         \includegraphics[width=\textwidth]{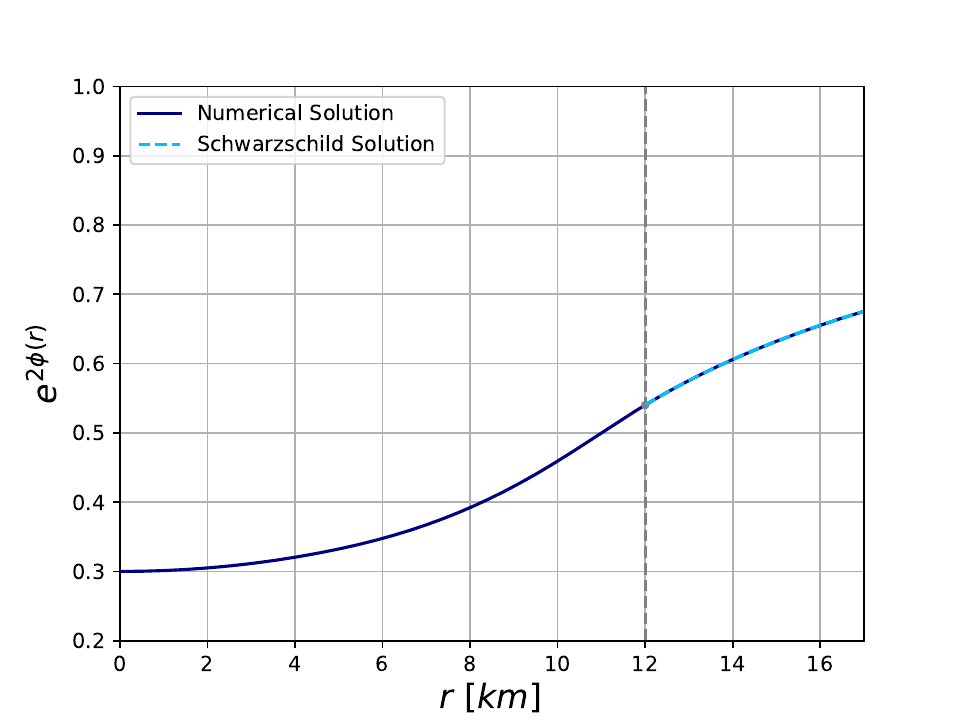}
         \caption{Metric function $e^{2\phi(r)}$ in model 4.}
         \label{phi_c2}
     \end{subfigure}
\caption{Metric function $e^{2\phi(r)}$ for the two models presented in Table~\ref{CasesNS}. The radius of the star is represented by the vertical gray line.}
\label{A0NS}
\end{figure}

Finally, the metric functions $A(r)=e^{2\psi(r)}$ and $e^{\phi(r)}$ are shown in Figs.~\ref{psi0NS} and~\ref{A0NS}. Considering the metric function $A(r)$, we observe that both cases satisfy the Buchdahl condition given in~\eqref{acond}. Additionally, both metric functions correctly match the Schwarzschild solution beyond the star radius.

\subsection{Mass-to-radius relation}

\begin{figure}[h]
     \centering
     \begin{subfigure}[b]{0.49\textwidth}
         \centering
         \includegraphics[width=\textwidth]{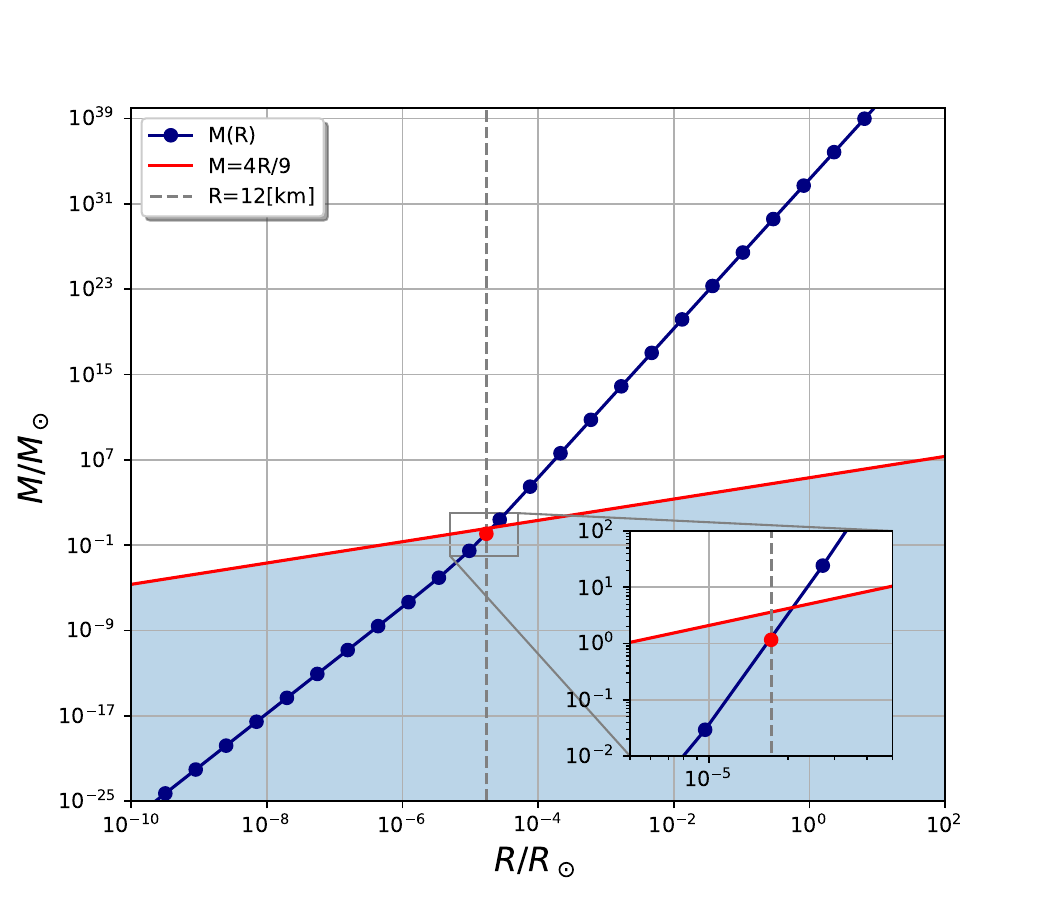}
         \caption{Mass-to-radius relation for model 3.}
         \label{MR_NS1}
     \end{subfigure}
     \hfill
     \begin{subfigure}[b]{0.49\textwidth}
         \centering
         \includegraphics[width=\textwidth]{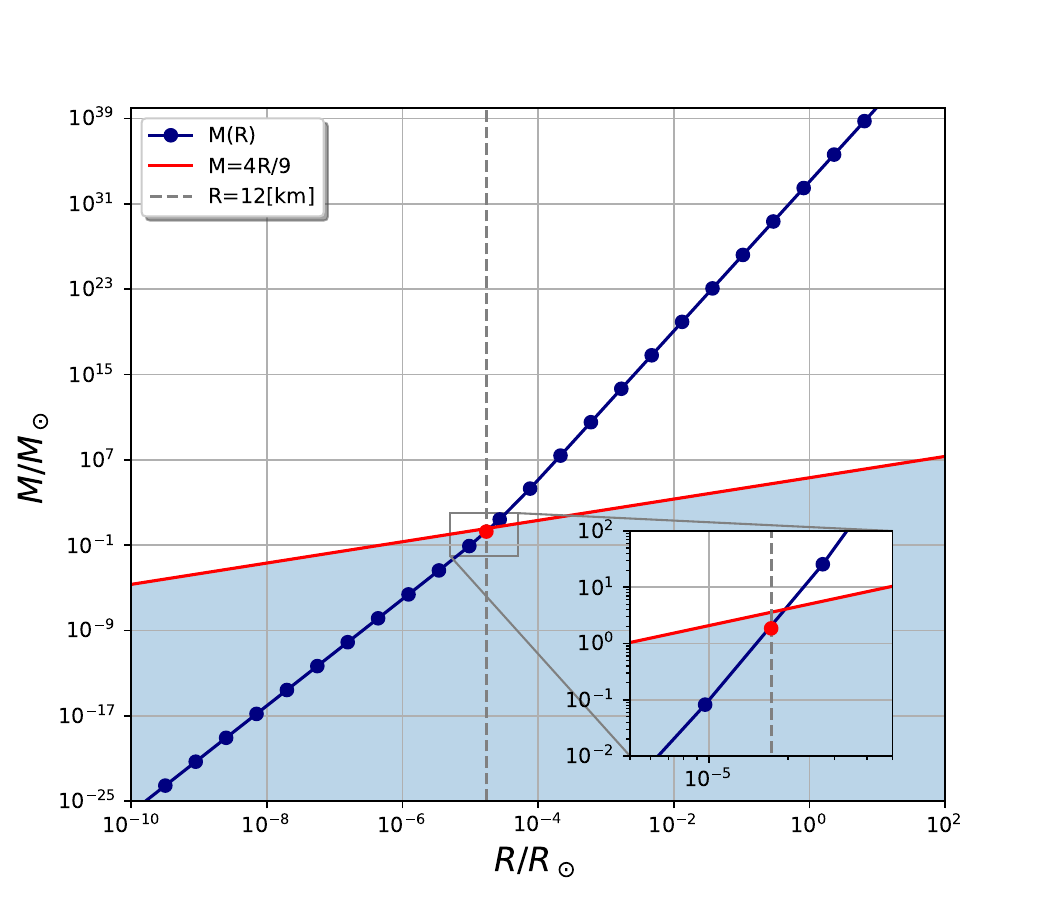}
         \caption{Mass-to-radius relation for model 4.}
         \label{MR_NS2}
     \end{subfigure}
    \caption{Relation between the total mass $M$ and radius $R$ as given by Eq.~\eqref{Mass29}. The parameters $\rho_c$ and $c_2$ are fixed as presented in Table~\ref{CasesNS}, while the parameter $c_4$ is allowed to vary. The points corresponding to the specific models of Table~\ref{CasesNS}, with all three parameters fixed, are highlighted as red points. The red solid lines correspond to the limiting case $M=R/9$; points below this line satisfy the Buchdahl limit.}
    \label{MR_NS}
\end{figure}

The mass-to-radius relationship for the model presented here can be obtained analytically from Eq.\eqref{mass} and is given by
\begin{equation}\label{Mass28}
M = 4\pi R^3 \left[\frac{\rho_c}{3}-\frac{c_2R^2}{5}-\frac{c_4R^4}{7}\right] \; .
\end{equation}
The explicit dependence of this polynomial function on the values of the constants $\rho_c$, $c_2$, and $c_4$ can in principle be very intricate.  However, using Eq.\eqref{rho_cR} which relates these three parameters to the total radius $R$, this relationship becomes quite simple as we can obtain an alternative expression for the total mass $M$ in terms of the radius $R$:
\begin{equation}\label{Mass29}
M(R) = \frac{8\pi R^3}{7} \left[ \frac{2}{3} \rho_c - \frac{1}{5} c_2R^2 \right] \; .
\end{equation}
Therefore, by fixing the two parameters $\rho_c$ and $c_2$ parameter we can determine the total mass as a function of the radius $M(R)$. This relationship is shown in Fig.~\ref{MR_NS}, where we fix the parameters for $\rho_0$ and $c_2$ as in Table~\ref{CasesNS}, and allow the parameter $c_4$ to vary. The specific points corresponding to models 3 and 4 (with all three parameters fixed) are highlighted in red. Additionally, the figure also shows a shaded region where the cases that satisfy the Buchdahl limit are located. Notice, however, that since the configurations analyzed in this work can also be used as initial conditions for a gravitational collapse,  the unshaded region also represents possible solutions. 

The total mass $M$ of the two specific models discussed in the previous section is presented in different units in Table~\ref{totalMass}. In both cases, the masses are comparable to the solar mass, $M_\odot=1.48 \: \mathrm{km}$.

\begin{table}[h]
\centering
\begin{tabular}{cccc}
\textbf{Total Mass} &  $\qquad[\mathrm{km}]\qquad$ & $\qquad[\mathrm{kg}]\qquad$ & $\qquad[\mathrm{M}_\odot]\qquad$ \\
         \hline
         \hline
        Model 3 & $1.72\ $ & $2.31\times 10^{24}\ $ & $1.16$ \\
        Model 4 & $2.76\ $ & $3.72\times 10^{24}\ $ & $1.86$
\end{tabular}
\caption{Total mass in kilometers and kilograms for the two cases presented in Table \ref{CasesNS}. }
\label{totalMass}
\end{table}


\subsection{Dust-free models }

\begin{figure}
     \centering
     \begin{subfigure}[b]{0.49\textwidth}
         \centering
         \includegraphics[width=\textwidth]{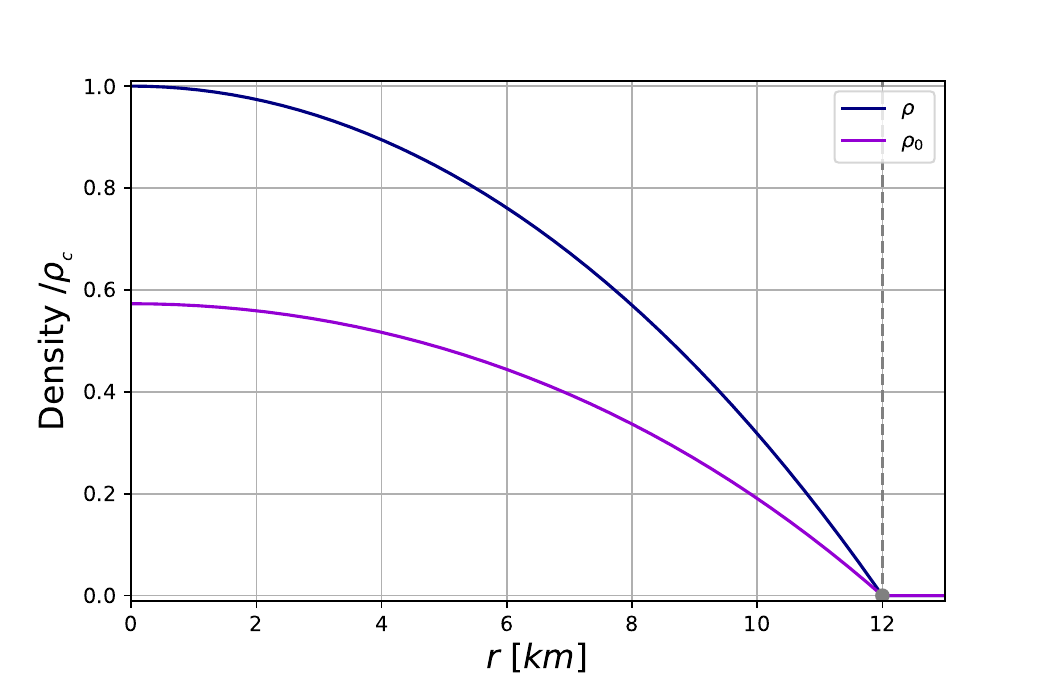}
         \caption{Total energy density for dust-free model.}
         \label{RhoSinPolvo}
     \end{subfigure}
     \hfill
     \begin{subfigure}[b]{0.49\textwidth}
         \centering
         \includegraphics[width=\textwidth]{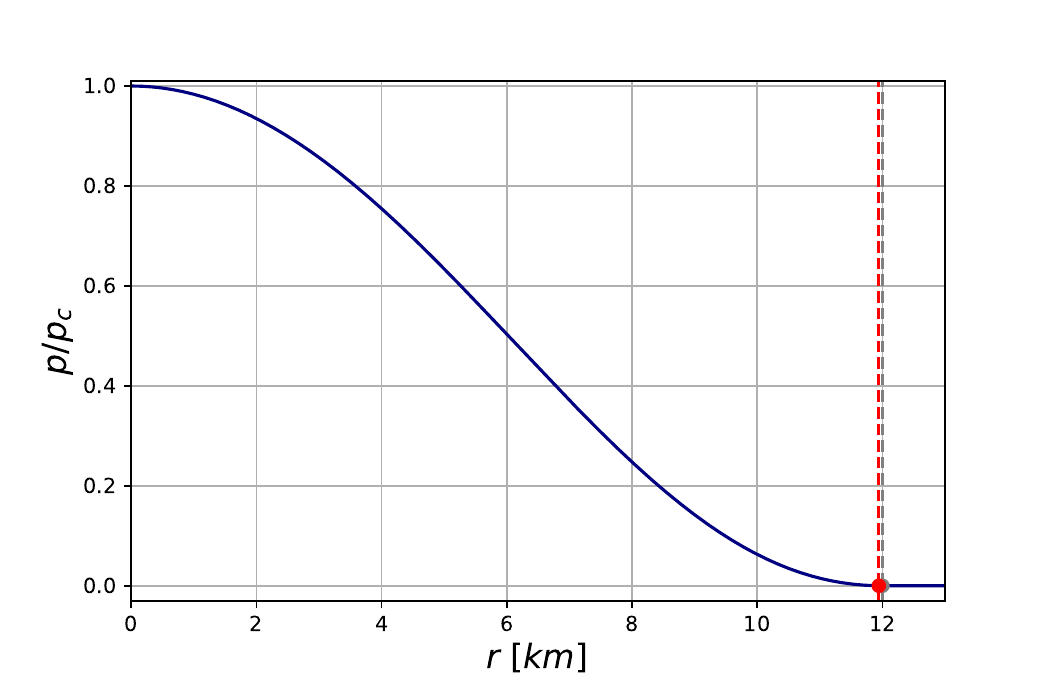}
         \caption{Pressure for dust-free model.}
         \label{pSinPolvo}
     \end{subfigure}
     \hfill
     \begin{subfigure}[b]{0.49\textwidth}
         \centering
         \includegraphics[width=\textwidth]{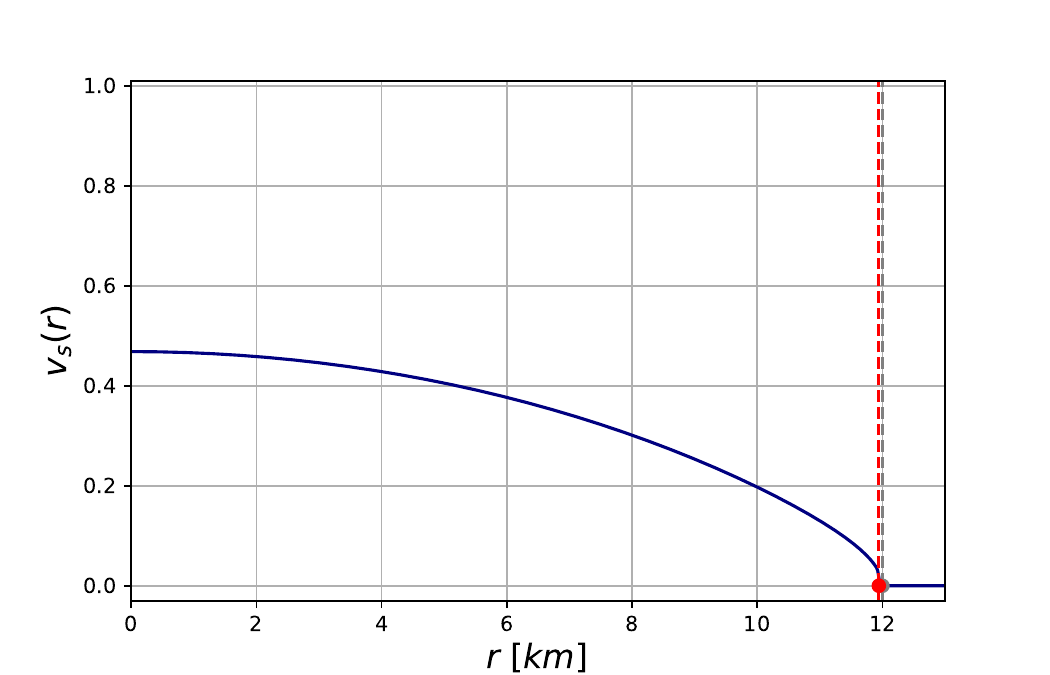}
         \caption{Sound speed for dust-free model.}
         \label{vsSinPolvo}
     \end{subfigure}
     \hfill
     \begin{subfigure}[b]{0.49\textwidth}
         \centering
         \includegraphics[width=\textwidth]{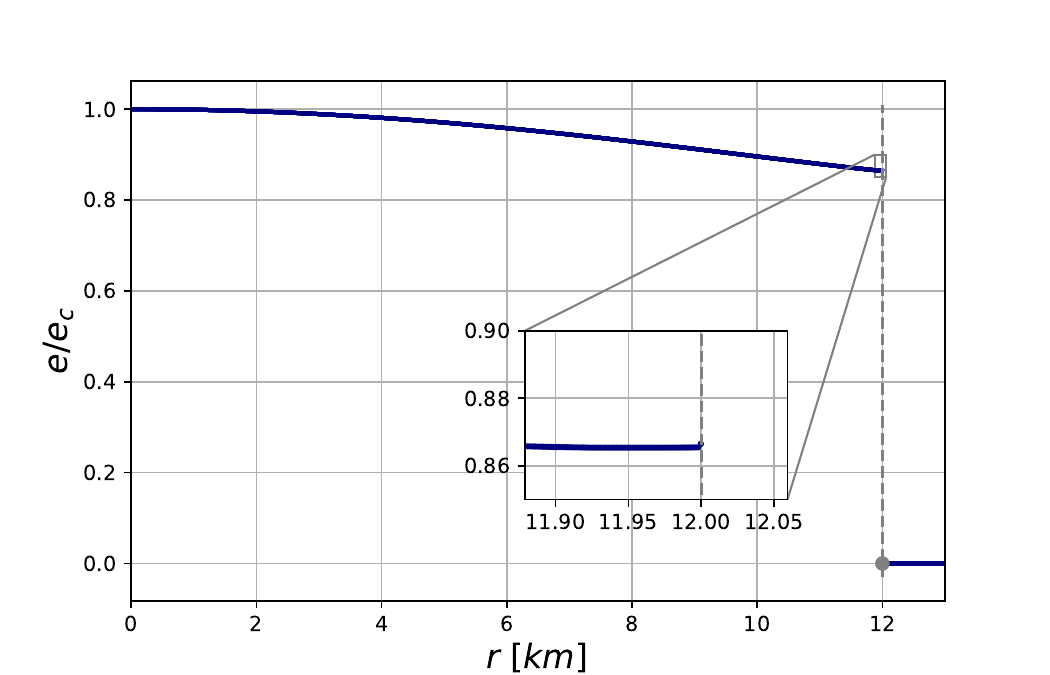}
         \caption{Specific internal energy for dust-free model.}
         \label{eSinPolvo}
     \end{subfigure}
     \hfill
     \begin{subfigure}[b]{0.49\textwidth}
         \centering
         \includegraphics[width=\textwidth]{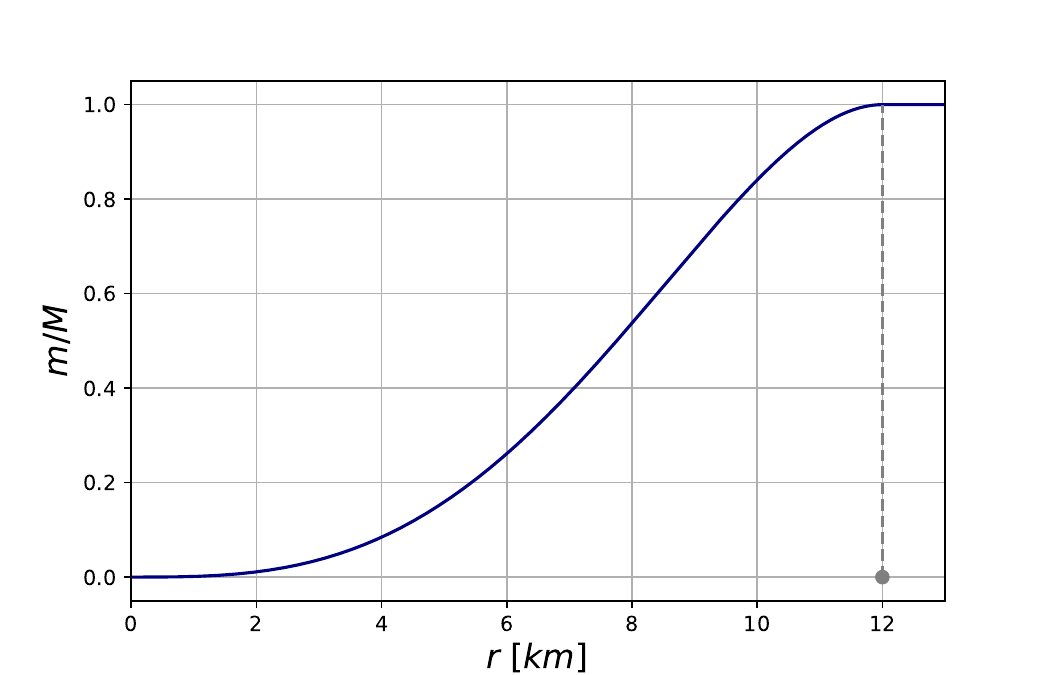}
         \caption{Mass function for dust-free model.}
         \label{mSinPolvo}
     \end{subfigure}
     \hfill
     \hfill
     \begin{subfigure}[b]{0.49\textwidth}
         \centering
         \includegraphics[width=\textwidth]{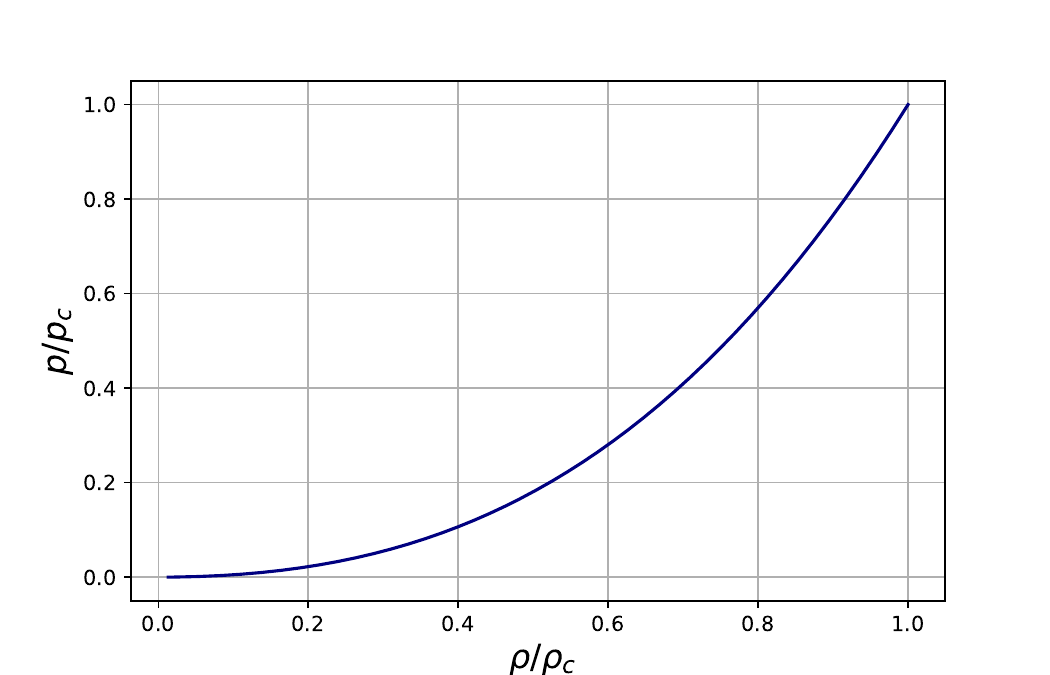}
         \caption{Equation of state for dust-free model.}
         \label{eosSinPolvo}
     \end{subfigure}
    \hfill
     \begin{subfigure}[b]{0.49\textwidth}
         \centering
         \includegraphics[width=\textwidth]{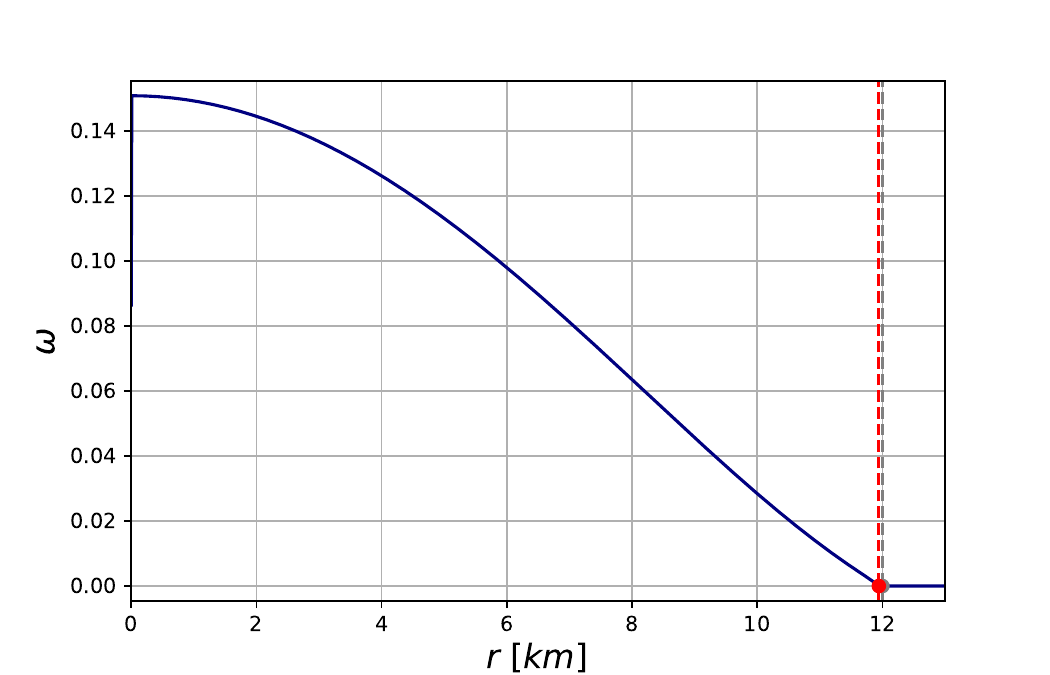}
         \caption{Barotropic function $\omega=p/\rho$ for dust-free model.}
         \label{omegaSinPolvo}
     \end{subfigure}
        \caption{Dust-free neutron star model.}
        \label{NS_SinPolvo}
\end{figure}

\begin{figure}[h]
     \centering
     \begin{subfigure}[b]{0.49\textwidth}
         \centering
         \includegraphics[width=\textwidth]{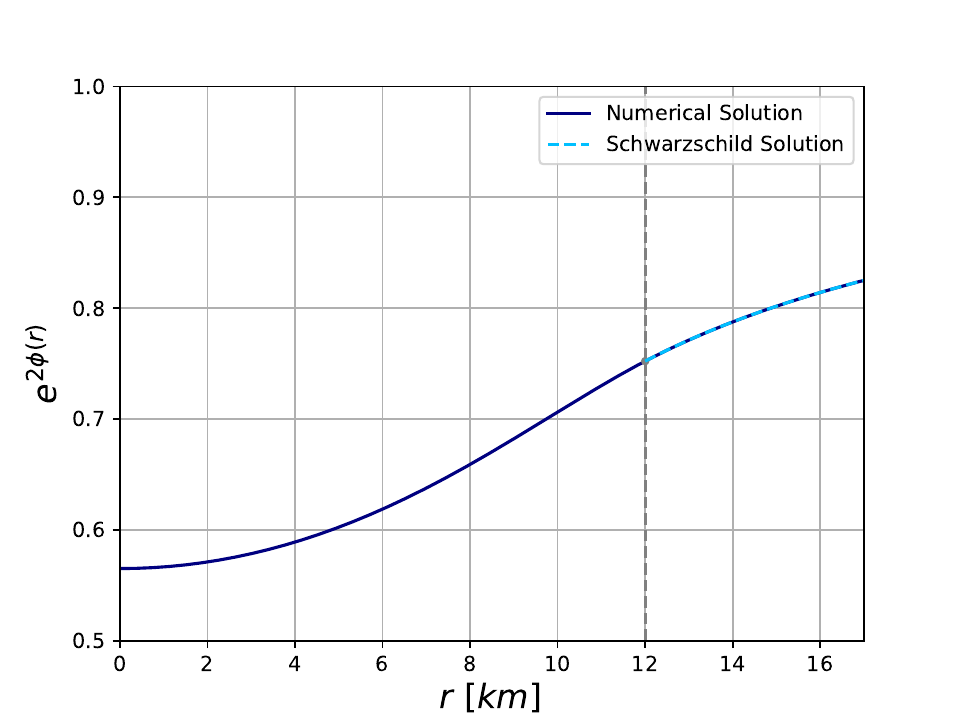}
         \caption{Metric function $e^{2\phi(r)}$ for dust-free model.}
         \label{psi0caso077DF}
     \end{subfigure}
     \hfill
     \begin{subfigure}[b]{0.5\textwidth}
         \centering
         \includegraphics[width=\textwidth]{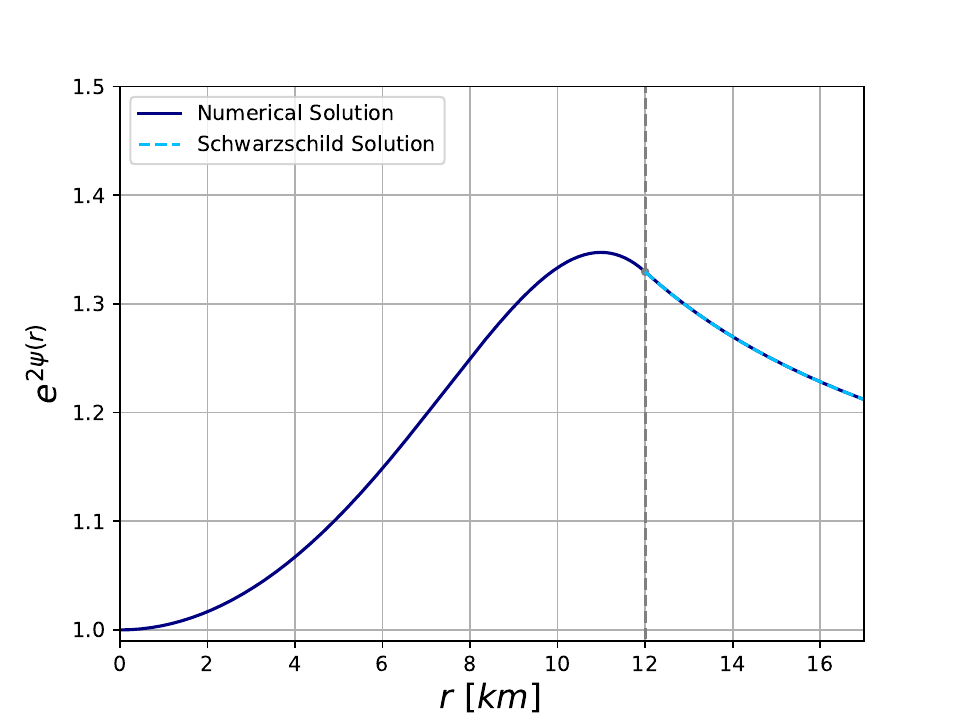}
         \caption{Metric function $e^{2\psi(r)}$ for dust-free model.}
         \label{psi0caso095DF}
     \end{subfigure}
        \caption{Metric functions $A(r)=e^{2\phi(r)}$ and $e^{2\psi(r)}$ for the dust-free model. The radius of the object is shown in gray.}
        \label{psi0DF}
\end{figure}

We will show here that it is possible to find limiting models with no dust layer. 
To this end, we keep the following three parameters fixed:  the radius of the object $R=12 \: \mathrm{km}$, the central rest energy density $\rho_{c}=2.87 \times 10^{-4} \: \mathrm{km}^{-2}$, and the central pressure is $p_c = 4.33\times 10^{-5} \: \mathrm{km}^{-2}$ (these values come from the examples of the previous section). Then, we choose the values of the parameters $c_2$ and $c_4$ for which the thickness of the dust layer tends to zero. This is reached for the values:
\begin{equation}
c_2 = 3.27 \times 10^{-6} \; , \qquad\qquad c_4 = 1.42\times 10^{-9} \; . 
\end{equation}
The resulting solutions are shown in Figs.~\ref{NS_SinPolvo} and~\ref{psi0DF}. Both the total and rest mass energy densities, as well as the pressure, decrease monotonically. The pressure reaches zero at a radius $r=11.95 \: \mathrm{km}$, resulting in the thinnest possible dust layer with a thickness of $0.05$. The behavior of the sound speed, as shown in Figure \ref{vsSinPolvo}, has a central and maximum value of $v_s=0.47$. In general, we see that all the functions show a behavior analogous to the plots presented in the previous sections. This shows that the presence of a dust layer around the central core facilitates the numerical integration of the model, but it is not necessary for the construction.

\section{Stability and convergence}
\label{sec:StaCon}

\begin{figure}[H]
     \centering
     \begin{subfigure}[b]{0.49\textwidth}
         \centering
         \includegraphics[width=\textwidth]{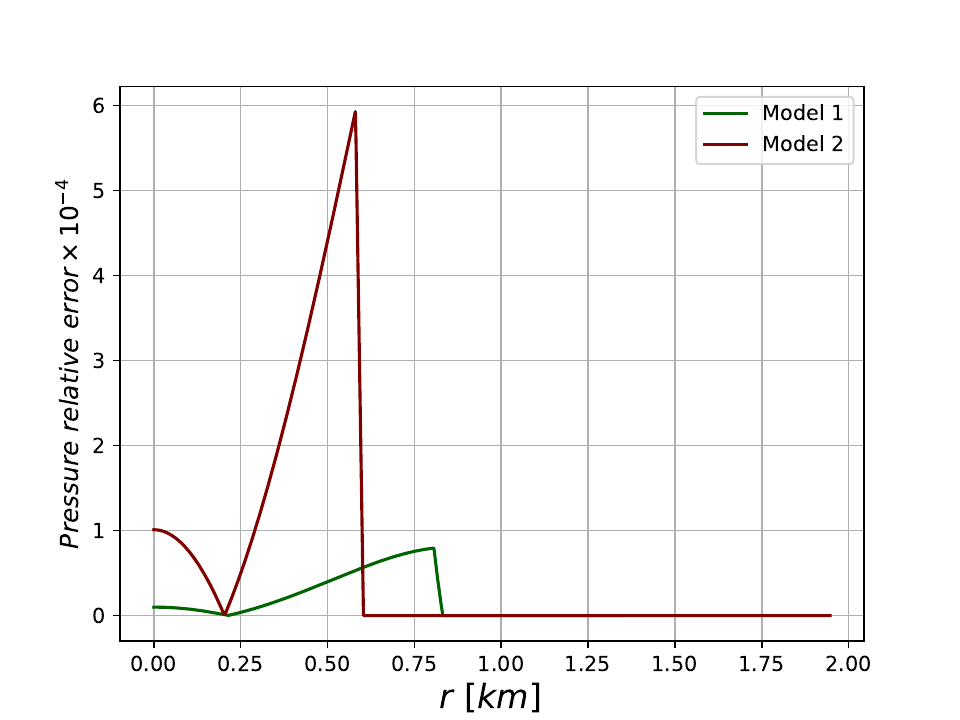}
         \caption{Relative error of the pressure $p$ for the models of the section \ref{sec:numSol}.}
         \label{rePressure}
     \end{subfigure}
     \hfill
     \begin{subfigure}[b]{0.49\textwidth}
         \centering
         \includegraphics[width=\textwidth]{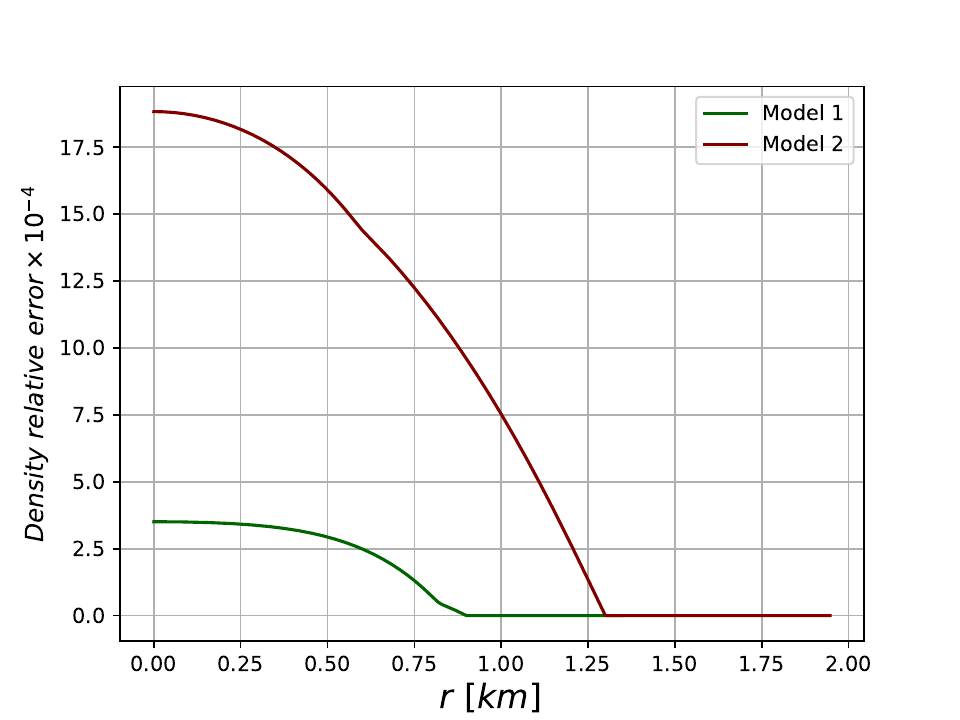}
         \caption{Relative error of the rest energy density $\rho_0$ for the models of the section \ref{sec:numSol}.}
         \label{reDensity}
     \end{subfigure}
        \caption{Relative error for the models of section \ref{sec:numSol} given in Table~\ref{Cases} when the parameters are modified by $0.5\%$.}
        \label{relError}
\end{figure}

\begin{figure}[H]
     \centering
     \begin{subfigure}[b]{0.49\textwidth}
         \centering
         \includegraphics[width=\textwidth]{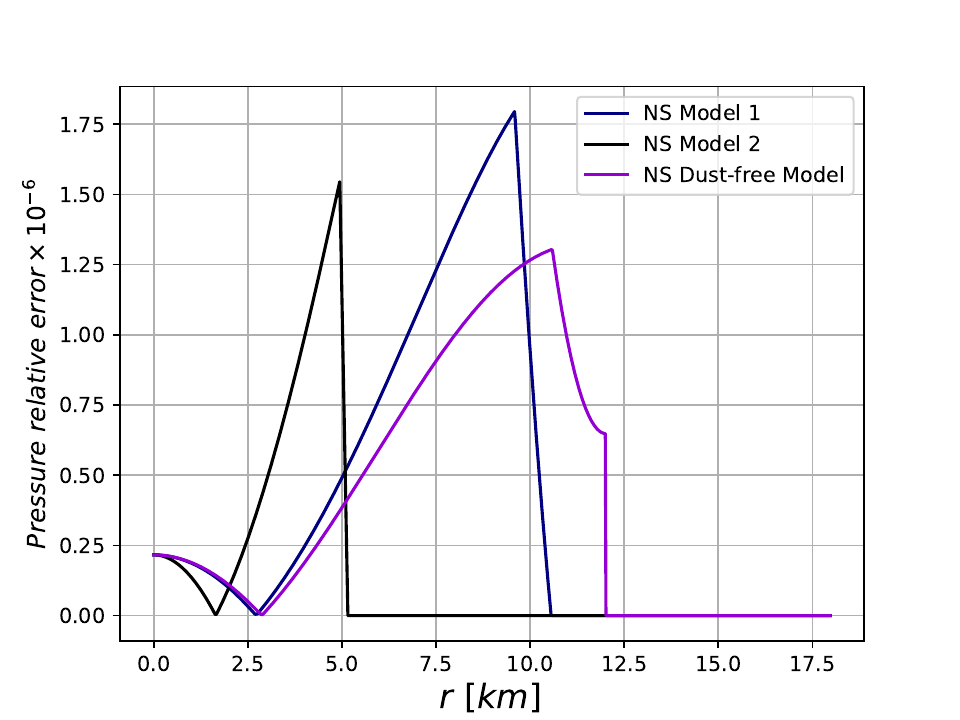}
         \caption{Relative error of the pressure for the neutron star cases.}
         \label{rePressureNS}
     \end{subfigure}
     \hfill
     \begin{subfigure}[b]{0.49\textwidth}
         \centering
         \includegraphics[width=\textwidth]{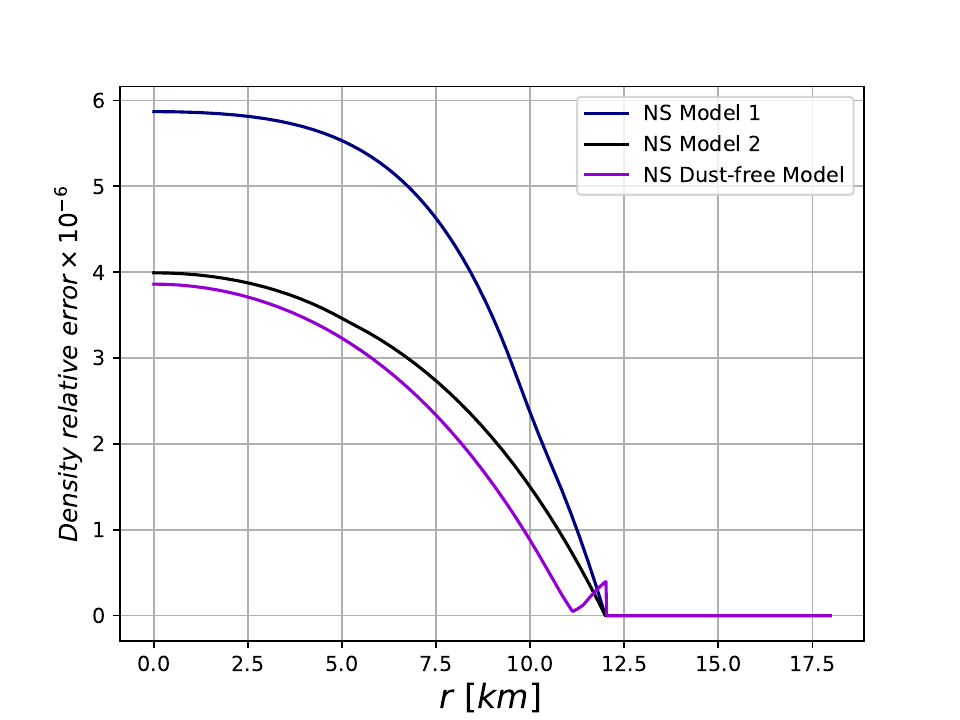}
         \caption{Relative error of the rest mass energy density for the neutron star cases.}
         \label{reDensityNS}
     \end{subfigure}
        \caption{Relative error for the neutron star cases given in Table~\ref{CasesNS} when the parameters are modified by $0.5\%$.}
        \label{relErrorNS}
\end{figure}

To test the stability of the model presented here, we varied the five parameters given in Tables  \ref{Cases} and \ref{CasesNS} by a factor of $0.5\%$. The resulting relative errors are presented in Figs.~\ref{relError} and~\ref{relErrorNS} for the models of section \ref{sec:numSol} and the neutron star models, respectively. We observe that the maximum relative error in the models of the section \ref{sec:numSol} is of the order of $10^{-3}$ specifically for the rest energy density. This indicates that the algorithm maintains high accuracy under small perturbations of the parameters, with the rest mass energy density being the most sensitive quantity to these changes.

To study the convergence of the numerical solutions the Hamiltonian constraint is a frequently used tool as shown, for example, in~\cite{alcubierre2011formulations}. However, in our case, this is satisfied exactly, as its expression  
\begin{equation}
\frac{dA}{dr} = A\left[\frac{1-A}{r} + 8\pi rA\rho\right] \; , \label{dAdr}
\end{equation}
can be derived without the numerical solution. This is because all functions are fixed analytically. The metric function $A(r)$ is given by the expression~ \eqref{Ar} and the total energy density is given by~\eqref{2Rho}.  In fact, the demonstration is immediate and we find
\begin{equation}
\der{A}{r} =
\begin{cases}
16 \pi r A(r)^2 \left[ \frac{1}{3}\rho_c - \frac{2}{5}c_2r^2 - \frac{3}{7}c_4r^4 \right] \; , & r<R \\
- \frac{2M}{(r-2M)^2} \; , & r>R\ .
\end{cases}
\end{equation}

\begin{figure}[H]
     \centering
     \begin{subfigure}[b]{0.49\textwidth}
         \centering
         \includegraphics[width=\textwidth]{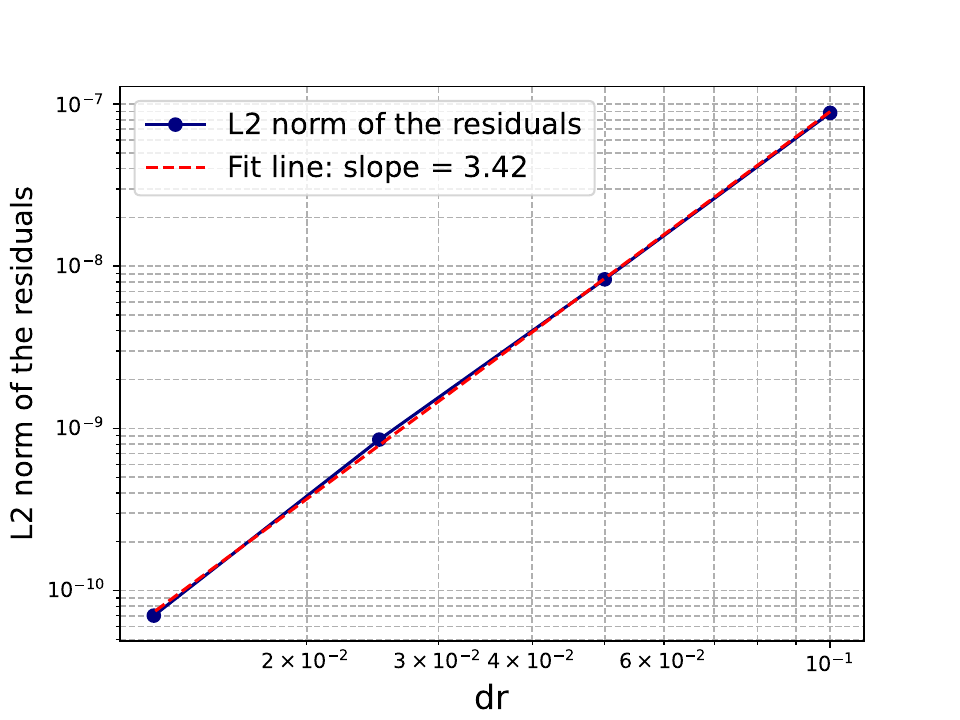}
         \caption{Convergence for model 1 of section \ref{sec:numSol}.}
         \label{L2C1}
     \end{subfigure}
     \hfill
     \begin{subfigure}[b]{0.49\textwidth}
         \centering
         \includegraphics[width=\textwidth]{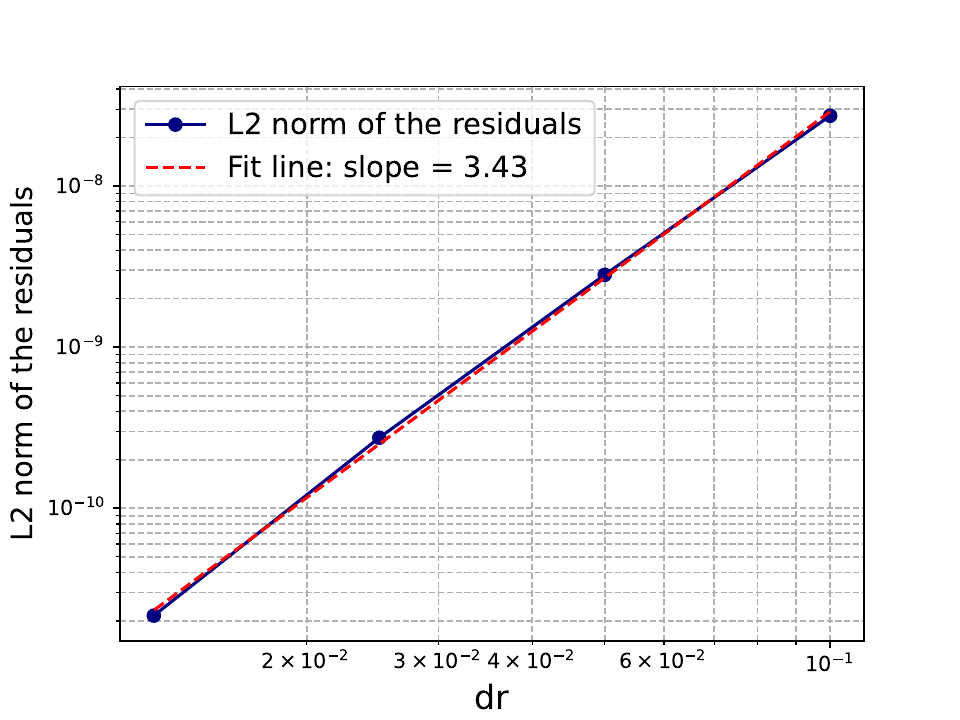}
         \caption{Convergence for model 2 the section \ref{sec:numSol}.}
         \label{L2C2}
     \end{subfigure}
        \caption{Convergence of the RK4 method for the models of the section \ref{sec:numSol}. The $L_2$ norm of the difference between a reference solution and the solutions obtained by systematically reducing the step size $\Delta r$ is plotted on a logarithmic scale. The slope must be four for a code of fourth order. }
        \label{L2}
\end{figure}

\begin{figure}[H]
     \centering
     \begin{subfigure}[b]{0.49\textwidth}
         \centering
         \includegraphics[width=\textwidth]{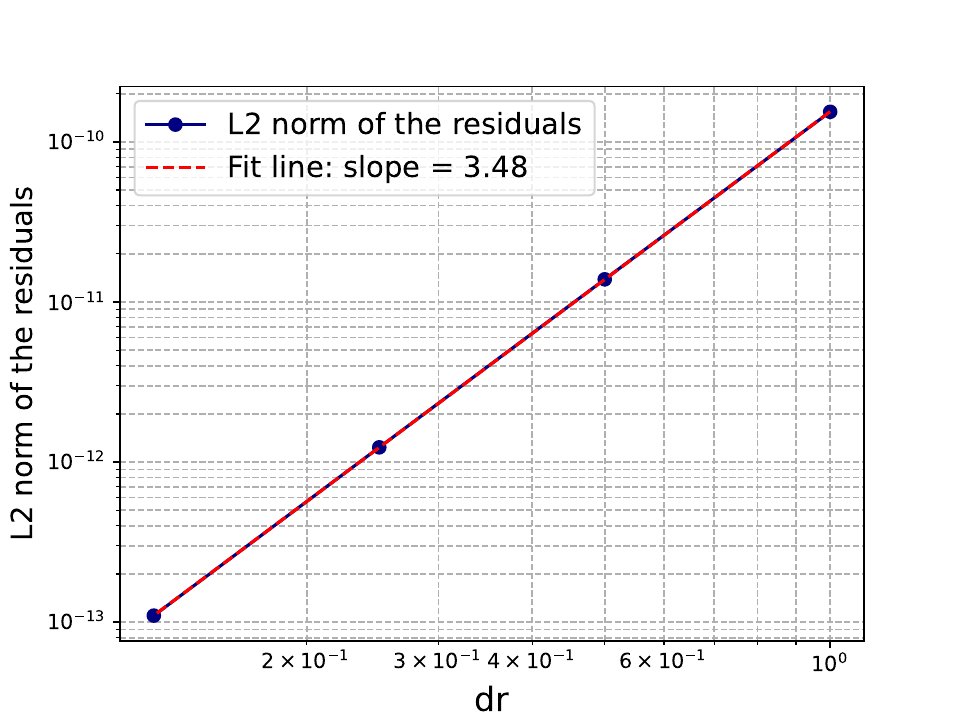}
         \caption{Convergence of the model 1 for neutron stars.}
         \label{L2C1_ns}
     \end{subfigure}
     \hfill
     \begin{subfigure}[b]{0.49\textwidth}
         \centering
         \includegraphics[width=\textwidth]{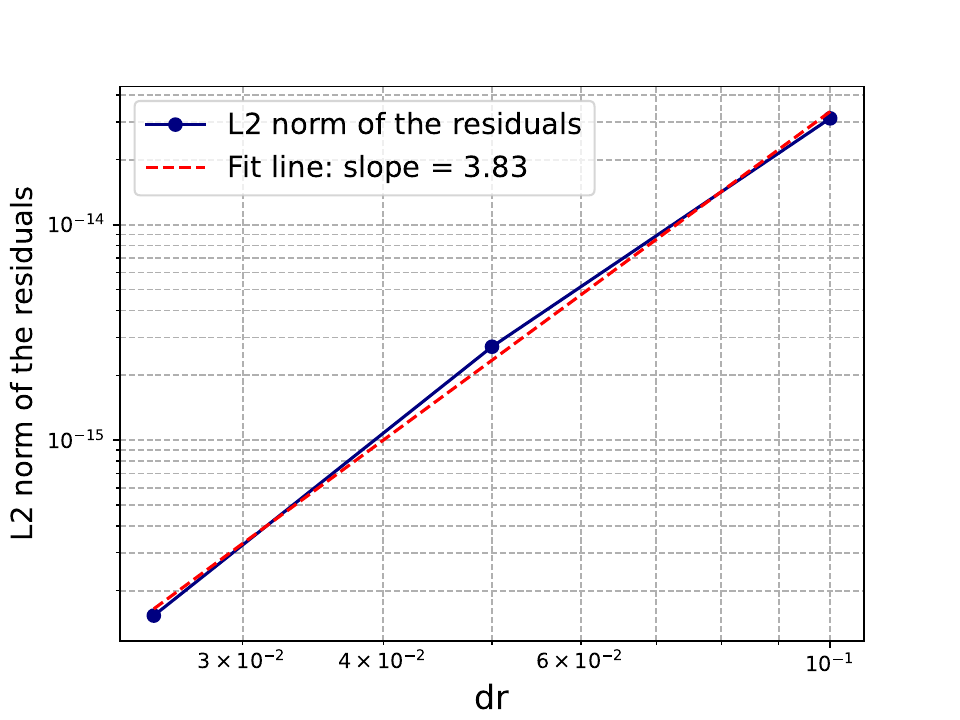}
         \caption{Convergence of the model 2 for neutron stars.}
         \label{L2C2_ns}
     \end{subfigure}
        \caption{Convergence of RK4 method for the neutron star cases. The $L_2$ norm of the difference between a reference solution and the solutions obtained by systematically reducing the step size $\Delta r$ is plotted on a logarithmic scale. The slope must be four for a code of fourth order.}
        \label{L2_ns}
\end{figure}

Therefore, in order to study the convergence we use the standard numerical. Specifically, we varied the value of the step size $\Delta r$ and observed the resulting convergence order using a reference solution for the pressure with $\Delta r_\mathrm{ref} = 1 \times 10^{-5}$. The numerical method (RK4) is of fourth order; thus, the error in the pressure should decrease by a factor of $2^{4}=16$, when the step size is halved. We calculate the $L_2$ norm of the difference between the reference solution and the solutions obtained by systematically reducing the mesh size. As a result, we obtain the set of plots shown in Figs.~\ref{L2} and \ref{L2_ns}, which show explicitly the convergence of our numerical solutions.

\section{Conclusions}
\label{sec:con}

In this work, we investigated spherically symmetric solutions of the Einstein equations with a perfect fluid as the source of gravity. Although there are many solutions in the literature with these 
characteristics, we focus on solutions that could be used to describe the gravitational field of realistic mass distributions. To this end, we demand that the solutions are free of singularities inside the fluid, present positive energy and pressure profiles, are in accordance with the Buchdahl limit, and are such that the sound speed satisfies the light speed limit.

To solve the above problem, an EoS is usually given {\it a priori}\/ that is commonly taken as a barotropic or a polytropic EoS. Our approach is different. We begin with a predetermined energy density profile, which is given as a polynomial function of the radial coordinate with arbitrary coefficients, that can be fixed in such a way as to guarantee that the energy density is consistent with the physical expectations. In particular, we study profiles in which the energy is a decreasing function of the radial distance with its maximum at the center and its minimum (vanishing density) at the radius of the star. We show that the properties of the model are essentially determined by a set of parameters that correspond to the values of the energy density and pressure at the center of the object and two constants that determine the energy density profile. 

We solve the TOV equations to determine the pressure profile and the metric functions, imposing the physical conditions mentioned above. It turns out that the configurations that agree with all physical requirements consist essentially of a central core surrounded by a layer of dust, whose thickness depends on the values chosen for the central pressure and the energy density profile. First, we test our model in two different cases, in which the values of the parameters determining the model are chosen arbitrarily within the intervals that are allowed by the physical requirements. As a second example, we consider two cases with parameter values fixed by the physical properties of canonical neutron stars. This shows the physical consistency of our model.  In addition, we tested the stability and convergence properties of our numerical solutions by using standard criteria for numerical methods. As for the stability test, we varied the five parameters given in Tables~\ref{Cases} and~\ref{CasesNS} by a factor of $0.5\%$ and found that the maximum relative error is of the order of $10^{-3}$ for the rest mass energy density. We conclude that the algorithm maintains high accuracy under small changes in the parameters.

To study the convergence we varied the step size of the numerical mesh $\Delta r$ and analyzed the resulting convergence order for the pressure using a reference solution with $\Delta r_\mathrm{ref} = 1 \times 10^{-4}$. The numerical method RK4 has an expected convergence factor of $2^{4}=16$ when the step size is halved, as we observe when we calculate the $L_2$ norm of the difference between the reference solution and the solutions obtained by systematically reducing the mesh width. This proves explicitly the convergence of our approach.

The model presented here can be used as the initial state for the study of the gravitational collapse of spherically symmetric mass distribution by using the tools of numerical relativity. Since the behavior of the physical quantities involved in the model can be controlled by choosing the parameters of the perfect fluid and pressure, we expect that a similar procedure could be used to control the dynamical evolution of the collapse. In particular, the dust layer, which surrounds the central core of the object in our model, seems to be an appropriate initial state for gravitational collapse since the dust should collapse under the sole action of the gravitational field of the central core. We expect to study the dynamical collapse of our models in future works.


\acknowledgments

This work was partially supported by CONACyT Network Projects No. 376127 and No. 304001, and DGAPA-UNAM project IN100523. The work of HQ was supported by UNAM PASPA-DGAPA.


\bibliography{0bibliografy}


\end{document}